\pdfoutput=1

\documentclass[camera,letterpaper,nomarginnotes,nonarrowgutter]{jpaper}

\usepackage[normalem]{ulem}
\usepackage{subcaption}
\usepackage{clipboard}
\usepackage{soul}
\usepackage{tikz}
\usepackage{flushend}
\usepackage[noadjust]{cite}
\usepackage{xargs} 
\usepackage{mathtools}
\usepackage{enumitem}
\usepackage{booktabs}
\usepackage{multirow}
\usepackage[many]{tcolorbox}
\usepackage{longfbox}

\usepackage{algorithmic}
\usepackage{graphicx}
\usepackage{textcomp}
\usepackage{xcolor}
\usepackage{pifont}
\usepackage[acronym,nonumberlist,nowarn]{glossaries}
    \glsdisablehyper
     \newacronym{ADI}{ADI}{Alternating Direction Implicit}
\newacronym{AGU}{AGU}{Address Generation Unit}
\newacronym[longplural={Access History Tables}]{AHT}{AHT}{Access History Table}
\newacronym{AHT-R}{AHT-R}{Access History Table - Read}
\newacronym{AHT-W}{AHT-W}{Access History Table - Write-Back}
\newacronym{AIP}{AIP}{Access Interval Predictor}
\newacronym{AL}{AL}{Added Latency for column accesses}
\newacronym{ASIC}{ASIC}{Application Specific Integrated Circuit}
\newacronym{AVX}{AVX}{Advanced Vector Extensions}
\newacronym[longplural={Basic Block Vectors}]{BBV}{BBV}{Basic Block Vector}
\newacronym{BHT}{BHT}{Branch History Table}
\newacronym{HBM}{HBM}{Hybrid Bandwidth Memory}
\newacronym{DDR4}{DDR4}{Double-Data Rate 4}
\newacronym{MIMD}{MIMD}{multiple-instruction multiple-data}
\newacronym{SIMT}{SIMT}{Single Instruction Multiple Threads}
\newacronym{SLP}{SALP}{subarray-level parallelism}
\newacronym{BLP}{BLP}{bank-level parallelism}

\newacronym{NN}{NN}{neural network}
\newacronym{BTB}{BTB}{Branch Target Buffer}
\newacronym{CAS}{CAS}{Column Address Strobe}
\newacronym{AMAT}{AMAT}{Average Memory Access Time}
\newacronym{CCD}{CCD}{Column to Column Delay}
\newacronym{AI}{AI}{Arithmetic Intensity}
\newacronym[longplural={Columnar Database Systems}]{CDBS}{CDBS}{Columnar Database System}
\newacronym[longplural={Chip Multiprocessors}]{CMP}{CMP}{Chip Multiprocessor}
\newacronym{CMT}{CMT}{Chip Multithreading}
\newacronym[longplural={Coarse-Grain Reconfigurable Arrays}]{CGRA}{CGRA}{Coarse-Grain Reconfigurable Array}
\newacronym{C-RAM}{C-RAM}{Computational-RAM}
\newacronym{CWD}{CWD}{Column Write Delay}
\newacronym{DBI}{DBI}{Dynamic Binary Instrumentation}
\newacronym[longplural={Database Management Systems}]{DBMS}{DBMS}{Database Management System}
\newacronym{DDR}{DDR}{Double Data Rate}
\newacronym{DEWP}{DEWP}{Dead Line and Early Write-Back Predictor}
\newacronym{DRAM}{DRAM}{Dynamic Random Access Memory}
\newacronym{DSBP}{DSBP}{Dead Sub-Block Predictor}
\newacronym{DSM}{DSM}{Decomposed Storage Manage}
\newacronym{FAW}{FAW}{Four row Activation Window}
\newacronym{FP}{FP}{Floating-Point}
\newacronym{EDP}{EDP}{Energy-Delay Product}
\newacronym{FCFS}{FCFS}{First-Come First-Serve}
\newacronym{FIFO}{FIFO}{Fist In, First Out}
\newacronym{FSM}{FSM}{finite state machine}
\newacronym{FPGA}{FPGA}{Field-Programmable Gate Array}
\newacronym[longplural={Functional Units}]{FU}{FU}{Functional Unit}
\newacronym{GAg}{GAg}{Global Adaptive branch prediction using one Global PHT}
\newacronym{GAs}{GAs}{Global Adaptive branch prediction using per-Set PHT}
\newacronym{GCC}{GCC}{GNU Compiler Collection}
\newacronym{GEMS}{GEMS}{General Execution-driven Multiprocessor Simulator}
\newacronym[longplural={General Purpose Processors}]{GPP}{GPP}{General Purpose Processor}
\newacronym{HMC}{HMC}{Hybrid Memory Cube}
\newacronym{HIVE}{HIVE}{HMC Instruction Vector Extensions}
\newacronym{IATAC}{IATAC}{Inter-Access Time per Access Count}
\newacronym{ILP}{ILP}{Instruction Level Parallelism}
\newacronym{IPC}{IPC}{Instructions per Cycle}
\newacronym{ISA}{ISA}{Instruction Set Architecture}
\newacronym{IRAM}{IRAM}{Intelligent RAM}
\newacronym{KIPS}{KIPS}{Kilo Instructions per Second}
\newacronym{LDS}{LDS}{Linked Data Structure}
\newacronym{LFSR}{LFSR}{Linear Feedback Shift Register}
\newacronym{LFMR}{LFMR}{Last-to-First Miss-Ratio}
\newacronym{MLP}{MLP}{Memory-Level Parallelism}

\newacronym{LLC}{LLC}{last-level cache}
\newacronym{LRU}{LRU}{Least Recently Used}
\newacronym{LSU}{LSU}{Load Store Unit}
\newacronym{LTP}{LTP}{Last-Touch Predictor}
\newacronym{LvP}{LvP}{Live-time Predictor}
\newacronym{LWP}{LWP}{Last Write Predictor}
\newacronym{MARSS}{MARSS}{Micro Architectural and System Simulator}
\newacronym{McPAT}{McPAT}{Multi-core Power, Area, and Timing}
\newacronym{MIPS}{MIPS}{Microprocessor without Interlocked Pipeline Stages}
\newacronym{MOB}{MOB}{Memory Order Buffer}
\newacronym{MMU}{MMU}{memory management unit}
\newacronym{PuD}{PUD}{processing-using-DRAM}
\newacronym{MPKI}{MPKI}{Misses per Kilo-Instruction}
\newacronym{MSHR}{MSHR}{Miss-Status Handling Registers}
\newacronym{NAS}{NAS}{Numerical Aerodynamic Simulation}
\newacronym{NDP}{NDP}{Near-Data Processing}
\newacronym{NMP}{NMP}{Near Memory Processor}
\newacronym{NoC}{NoC}{Network-on-Chip}
\newacronym{NPB}{NPB}{NAS Parallel Benchmark}
\newacronym{NUCA}{NUCA}{Non-Uniform Cache Architecture}
\newacronym{NUMA}{NUMA}{Non-Uniform Memory Access}
\newacronym{OoO}{OoO}{Out-of-Order}
\newacronym{OpenMP}{OpenMP}{Open Multi-Processing}
\newacronym{OS}{OS}{Operating System}
\newacronym{PAg}{PAg}{Per-address Adaptive branch prediction using one Global PHT}
\newacronym{PAs}{PAs}{Per-address Adaptive branch prediction using per-Set PHT}
\newacronym{PC}{PC}{Program Counter}
\newacronym[longplural={Pointer-Chasing Engines}]{PCE}{PCE}{Pointer-Chasing Engine}
\newacronym{PCM}{PCM}{Performance Counter Monitor}
\newacronym{PIM}{PIM}{processing-in-memory}
\newacronym[longplural={Pattern History Tables}]{PHT}{PHT}{Pattern History Table}
\newacronym{RAPL}{RAPL}{Running Average Power Limit}
\newacronym{GEMM}{GEMM}{general matrix-matrix multiplication}
\newacronym{RAS}{RAS}{Row Address Strobe}
\newacronym{RAT}{RAT}{Registers Alias Table}
\newacronym{RC}{RC}{Row Cycle}
\newacronym{RCD}{RCD}{RAS to CAS Delay}
\newacronym[longplural={Row-based Database Systems}]{RDBS}{RDBS}{Row-based Database System}
\newacronym{ROB}{ROB}{Reorder Buffer}
\newacronym{RRD}{RRD}{Row to Row activation Delay}
\newacronym{DNN}{DNN}{Deep Neural Network}
\newacronym{ANN}{ANN}{Artificial Neural Network}
\newacronym{PnM}{PNM}{processing-near-memory}
\newacronym{PuM}{PUM}{processing-using-memory}
\newacronym{FFD}{FFD}{first-fit-decreasing}
\newacronym{HFF}{HFF}{helper flip-flop}
\newacronym{OBPS}{OBPS}{one-bit per-subarray}
\newacronym{ABOS}{ABOS}{all-bits in one-subarray}
\newacronym{ABPS}{ABPS}{all-bits per-subarray}
\newacronym{GEMV}{GEMV}{general matrix-vector product}

\newacronym{RBR}{RBR}{redundant binary representation}
\newacronym{RP}{RP}{Row Precharge}
\newacronym{RTL}{RTL}{Register Transfer Level}
\newacronym{RTP}{RTP}{Read To Precharge}
\newacronym{RVU}{RVU}{Reconfigurable Vector Unit}
\newacronym{SDP}{SDP}{Skewed Dead-Block Predictor}
\newacronym{SESC}{SESC}{Superescalar Simulator}
\newacronym{SSE}{SSE}{Streaming SIMD Extensions}
\newacronym{SFP}{SFP}{Spatial Footprint Predictor}
\newacronym{SiNUCA}{SiNUCA}{Simulator of Non-Uniform Cache Architectures}
\newacronym{SMT}{SMT}{Simultaneous Multi-Threading}
\newacronym{SIMD}{SIMD}{single-instruction multiple-data}
\newacronym{LUT}{LUT}{Lookup Table}

\newacronym{SPEC}{SPEC}{Standard Performance Evaluation Corporation}
\newacronym{SoC}{SoC}{System-on-Chip}
\newacronym{SPP}{SPP}{Spatial Pattern Predictor}
\newacronym{SRAM}{SRAM}{Static Random Access Memory}
\newacronym{SSOR}{SSOR}{Symmetric Successive Over-Relaxation}
\newacronym{SSV}{SSV}{Search Set Vector}
\newacronym{TLB}{TLB}{Translation Look-aside Buffer}
\newacronym{TLP}{TLP}{Thread Level Parallelism}
\newacronym[longplural={Through-Silicon Vias}]{TSV}{TSV}{Through-Silicon Via}
\newacronym{VWQ}{VWQ}{Virtual Write Queue}
\newacronym{WTR}{WTR}{Write Recovery time}
\newacronym{WR}{WR}{Write To Read delay time}
\newacronym{TRA}{TRA}{triple row activation}

\usepackage{xspace}
\usepackage[binary-units=true]{siunitx}
\usepackage{todonotes}
\usepackage[us,12hr]{datetime}
\usepackage[en-GB, useregional=numeric]{datetime2}
\usepackage{balance}
\usepackage{dblfloatfix}

\usepackage{setspace}
\usepackage[export]{adjustbox}

\usepackage{fancyhdr}

\usepackage{cleveref}
\crefformat{section}{\S#2#1#3}
\crefformat{subsection}{\S#2#1#3}
\crefformat{subsubsection}{\S#2#1#3}

\definecolor{airforceblue}{rgb}{0.36, 0.54, 0.66}
\definecolor{dodgerblue}{rgb}{0.12, 0.56, 1.0}
\definecolor{brandeisblue}{rgb}{0.0, 0.44, 1.0}
\definecolor{brickred}{rgb}{0.8, 0.25, 0.33}
\definecolor{eggplant}{rgb}{0.38, 0.25, 0.32}
\definecolor{byzantium}{rgb}{0.44, 0.16, 0.39}
\definecolor{ddgreen}{rgb}{0.00, 0.50, 0.00}

\definecolor{mygreen}{rgb}{0,0.6,0}
\definecolor{mygray}{rgb}{0.5,0.5,0.5}
\definecolor{mymauve}{rgb}{0.58,0,0.82}

\definecolor{bluehl}{rgb}{0.8,0.874,1}
\definecolor{pinkhl}{rgb}{0.992156863,0.847058824,1}
\definecolor{macaroniandcheese}{rgb}{1.0, 0.74, 0.53}
\definecolor{mossgreen}{rgb}{0.68, 0.87, 0.68}
\definecolor{greenhl}{rgb}{0.835,0.996,0.939}
\definecolor{yellowhl}{rgb}{0.996,0.957,0.8}
\definecolor{palecerulean}{rgb}{0.61, 0.77, 0.89}
\definecolor{gray(x11gray)}{rgb}{0.75, 0.75, 0.75}
\definecolor{amethyst}{rgb}{0.6, 0.4, 0.8}
\definecolor{ao}{rgb}{0.0, 0.5, 0.0}
\definecolor{burntorange}{rgb}{0.8, 0.33, 0.0}

\definecolor{cadmiumorange}{rgb}{0.93, 0.53, 0.18}

\definecolor{frenchlilac}{rgb}{0.53, 0.38, 0.56}
\definecolor{heliotrope}{rgb}{0.87, 0.45, 1.0}
\definecolor{peridot}{rgb}{0.9, 0.89, 0.0}
\definecolor{saffron}{rgb}{0.96, 0.77, 0.19}
\definecolor{tuscanred}{rgb}{0.51, 0.21, 0.21}
\definecolor{uscgold}{rgb}{1.0, 0.8, 0.0}
\definecolor{tangerineyellow}{rgb}{1.0, 0.8, 0.0}
\definecolor{rufous}{rgb}{0.66, 0.11, 0.03}
\definecolor{safetyorange}{rgb}{1.0, 0.4, 0.0}

\newif\ifcameraready
\camerareadytrue

\ifcameraready
    \newcommand{\gfcr}[1]{\textcolor{black}{#1}}
    \newcommand{\gfcri}[1]{\textcolor{black}{#1}} 
    \newcommand{\gfcrii}[1]{\textcolor{black}{#1}} 
    \newcommand{\gfcriii}[1]{\textcolor{black}{#1}} 
    \newcommand{\gfcriv}[1]{\textcolor{black}{#1}} 

    \newcommand{\omcri}[1]{\textcolor{black}{#1}}
    \newcommand{\omcrii}[1]{\textcolor{black}{#1}}
    \newcommand{\omcriii}[1]{\textcolor{black}{#1}}
    \newcommand{\omcriv}[1]{\textcolor{black}{#1}}

\else 
    \newcommand{\gfcr}[1]{\textcolor{black}{#1}} 
    \newcommand{\gfcri}[1]{\textcolor{black}{#1}} 
    \newcommand{\gfcrii}[1]{\textcolor{black}{#1}} 
    \newcommand{\gfcriii}[1]{\textcolor{black}{#1}} 
    \newcommand{\gfcriv}[1]{\textcolor{blue}{#1}} 
    
    \newcommand{\omcri}[1]{\textcolor{black}{#1}}
    \newcommand{\omcrii}[1]{\textcolor{black}{#1}}
    \newcommand{\omcriii}[1]{\textcolor{black}{#1}}
    \newcommand{\omcriv}[1]{\textcolor{blue}{#1}}

\fi

\newcommand{\versionnum}[0]{4.0}

\definecolor{MidnightBlue}{rgb}{0.1, 0.1, 0.44}

\newcommand{\li}{(\textit{i})}
\newcommand{\lii}{(\textit{ii})}
\newcommand{\liii}{(\textit{iii})}
\newcommand{\liv}{(\textit{iv})}

\definecolor{blush}{rgb}{0.87, 0.36, 0.51}

\newcommand{\prop}{\textit{Proteus}\xspace}

\newcommand\bbop{\emph{bbop}\xspace}

\newcommand\uproglib{\emph{Parallelism-Aware \uprog Library}\xspace}
\newcommand\dynengine{\emph{Dynamic Bit-Precision Engine}\xspace}
\newcommand\uprogunit{\emph{\uprog Select Unit}\xspace}
\newcommand\bitprec{\emph{Bit-Precision Calculator Unit}\xspace}
\newcommand\prelut{\emph{Pre-Loaded Cost Model \glspl{LUT}}\xspace}

\newcommand\costmodel{\emph{Cost Model Logic}\xspace}
\newcommand\gbidx{\emph{\textmu{}Program\_addr}\xspace}
\newcommand\idx{\emph{\textmu{}Program\_id}\xspace}

\newcommand\aap{\texttt{AAP}/\texttt{AP}\xspace}
\newcommand\aaps{\texttt{AAP}s/\texttt{AP}s\xspace}

\newcommand\uprog{\textmu{}Program\xspace}
\newcommand\uprogs{\textmu{}Programs\xspace}

\newcommand{\paratitle}[1]{\vspace{4pt}\noindent\textbf{#1.}}

\DeclarePairedDelimiter\ceil{\lceil}{\rceil}

\newcommand{\circledii}[1]{\tikz[baseline=(char.base)]{\node[shape=circle,draw,inner sep=0pt,fill=gray, text=white] (char) {\itshape#1};}}

\newcommand{\circlediii}[1]{\tikz[baseline=(char.base)]{\node[shape=circle,draw,inner sep=0pt,fill=white, text=black] (char) {\itshape#1};}}

\newcommand{\tempcommand}[1]{\renewcommand{\arraystretch}{#1}}

\newcommand\ignore[1]{ }
\newcommand{\revdel}[1]{}
\newcommand{\sgdel}[1]{}

\newcommand{\prtagA}[1]{\lfbox[padding=1pt, border-color=red, background-color=red!20]{\revA{\textbf{\scriptsize #1}}}}

\newcommand{\prtagD}[1]{\lfbox[padding=1pt, border-color=heliotrope, background-color=heliotrope!20]{\revD{\textbf{\scriptsize #1}}}}

\newif\ifasplosrevisionsubmission
\asplosrevisionsubmissiontrue

\ifasplosrevisionsubmission
    \newcommand{\omrev}[1]{#1}
\else
    \newcommand{\omrev}[1]{\textcolor{orange}{#1}}
\fi

\newif\ifasplosrevision
\asplosrevisionfalse
\ifasplosrevision 
    \newcommand{\asplosrev}[1]{\textcolor{blue}{#1}}
\else
    \newcommand{\asplosrev}[1]{\textcolor{black}{#1}}
    \renewcommand{\hl}[1]{#1}
\fi

\newif\ifasplossubmission
\asplossubmissiontrue
\ifasplossubmission 
    \newcommand{\gfasplos}[1]{\textcolor{black}{#1}}

    \newcommand{\agyasploscomment}[1]{}
\else
    \newcommand{\gfasplos}[1]{\textcolor{blue}{#1}}

    \newcommand{\agyasploscomment}[1]{\textcolor{red}{\textbf{!!!~Giray:} #1}}
    
\fi

\newif\ifmicrosubmission
\microsubmissiontrue
\ifmicrosubmission 
    \newcommand{\gfmicro}[1]{\textcolor{black}{#1}}
    \newcommand{\agymicro}[1]{\textcolor{black}{#1}}
    
    \newcommand{\agymicrocomment}[1]{}
\else
    \newcommand{\gfmicro}[1]{\textcolor{blue}{#1}}
    
    \newcommand{\agymicro}[1]{\textcolor{red}{#1}}
    \newcommand{\agymicrocomment}[1]{\textcolor{red}{\textbf{!!!~Giray:} #1}}
    
\fi

\newif\ifiscarevision
\iscarevisionfalse
\ifiscarevision 
    \newcommand{\revA}[1]{\textcolor{red}{#1}}
    \newcommand{\revB}[1]{\textcolor{mygreen}{#1}}
    \newcommand{\revC}[1]{\textcolor{safetyorange}{#1}}
    \newcommand{\revD}[1]{\textcolor{heliotrope}{#1}}
    \newcommand{\revE}[1]{\textcolor{rufous}{#1}}
    \newcommand{\revCommon}[1]{\textcolor{blue}{#1}}

    \newcommandx{\changeCM}[2][1=]{\todo[linecolor=blue,backgroundcolor=blue!25,bordercolor=blue,#1,size=\scriptsize]{\revCommon{\textbf{#2}}}}
    
    \newcommandx{\changeA}[2][1=]{\todo[linecolor=red,backgroundcolor=red!25,bordercolor=red,#1,size=\scriptsize]{\revA{\textbf{#2}}}}
    
    \newcommandx{\changeB}[2][1=]{\todo[linecolor=mygreen,backgroundcolor=mygreen!25,bordercolor=mygreen,#1,size=\scriptsize]{\revB{\textbf{#2}}}}
    
    \newcommandx{\changeC}[2][1=]{\todo[linecolor=safetyorange,backgroundcolor=safetyorange!25,bordercolor=safetyorange,#1,size=\scriptsize]{\revC{\textbf{#2}}}}
    
    \newcommandx{\changeD}[2][1=]{\todo[linecolor=heliotrope,backgroundcolor=heliotrope!25,bordercolor=heliotrope,#1,size=\scriptsize]{\revD{\textbf{#2}}}}
    
    \newcommandx{\changeE}[2][1=]{\todo[linecolor=rufous,backgroundcolor=rufous!25,bordercolor=rufous,#1,size=\scriptsize]{\revE{\textbf{#2}}}}
\else
    \newcommand{\revA}[1]{\textcolor{black}{#1}}
    \newcommand{\revB}[1]{\textcolor{black}{#1}}
    \newcommand{\revC}[1]{\textcolor{black}{#1}}
    \newcommand{\revD}[1]{\textcolor{black}{#1}}
    \newcommand{\revE}[1]{\textcolor{black}{#1}}
    \newcommand{\revCommon}[1]{\textcolor{black}{#1}}

    \newcommandx{\changeCM}[2][1=]{\todo[disable,#1]{#2}}
    \newcommandx{\changeA}[2][1=]{\todo[disable,#1]{#2}}
    \newcommandx{\changeB}[2][1=]{\todo[disable,#1]{#2}}
    \newcommandx{\changeC}[2][1=]{\todo[disable,#1]{#2}}
    \newcommandx{\changeD}[2][1=]{\todo[disable,#1]{#2}}
    \newcommandx{\changeE}[2][1=]{\todo[disable,#1]{#2}}
\fi

\newif\ifcut
\cutfalse

\ifcut
   \newcommand{\gfcut}[1]{} 
\else
    \newcommand{\gfcut}[1]{\textcolor{red}{\sout{#1}}}
\fi

\newif\ifiscasubmission
\iscasubmissiontrue

\ifiscasubmission
    \newcommand{\gfisca}[1]{#1}
    \newcommand{\gfbisca}[1]{}
    \newcommand{\sg}[1]{#1}
    \newcommand{\sgi}[1]{#1}
    \newcommand{\om}[1]{#1}
\else
    \newcommand{\gfbisca}[1]{\textcolor{blue}{\textit{GF: #1}}}
    \newcommand{\gfisca}[1]{\textcolor{blue}{#1}}

    \newcommand{\sg}[1]{\textcolor{red}{#1}}
    \newcommand{\sgi}[1]{\textcolor{brickred}{#1}}

    \newcommand{\om}[1]{\textcolor{orange}{#1}}
\fi

\newif\ifsubmission
\submissiontrue

\ifsubmission
    
    \newcommand{\juan}[1]{#1}
    \newcommand{\gf}[1]{#1}
    \newcommand{\gfii}[1]{#1}
    
    \newcommand{\jgl}[1]{}

    \newcommand{\gfb}[1]{}
    \newcommand{\mayank}[1]{}
    
    \newcommand{\agy}[1]{#1}
    \newcommand{\agycomment}[1]{}
\else
    \newcommand{\jgl}[1]{\textcolor{brickred}{\textit{JGL: #1}}}
    \newcommand{\gfb}[1]{\textcolor{blue}{\textit{GF: #1}}}
    \newcommand{\todo}[1]{\textcolor{red}{\textbf{TODO: #1}}}
    \newcommand{\juan}[1]{\textcolor{brickred}{#1}}
    \newcommand{\mayank}[1]{\textcolor{green}{\textit{Mayank: #1}}}
    
    \newcommand{\gf}[1]{\textcolor{blue}{#1}}

    \newcommand{\gfii}[1]{\textcolor{red}{#1}}

    \newcommand{\agy}[1]{\textcolor{orange}{#1}}
    \newcommand{\agycomment}[1]{\agy{\textbf{[@gy:} #1\textbf{]}}}

\fi

\newcommand{\circled}[1]{\tikz[baseline=(char.base)]{\node[shape=circle,draw,inner sep=0pt,fill=black, text=white] (char) {#1};}}

\newcommand\pimdef{\cite{ghose.ibmjrd19, mutlu2020modern,deoliveira2021IEEE,pim-book,mutlu2019processing,mutlu2019enabling,mutlu2015research,mutlu2013memory,loh2013processing,Near-Data,stone1970logic,Miss_Mem_Wall_1996}\xspace}

\newcommand\pnm{\cite{farmahini2015nda,babarinsa2015jafar,devaux2019true,ghiasi2022genstore,gomez2021benchmarkingcut,gomezluna2021benchmarking,gomez2022benchmarking,syncron,singh2020nero,skhynixpim,ke2021near,giannoula2022sparsep,shin2018mcdram,cho2020mcdram,denzler2021casper,asghari2016chameleon,IRAM_Micro_1997,C_RAM_1999,CASES_MVX,Xi_2015,sun2021abc,matam2019graphssd,gokhale1995processing,hall1999mapping,MEMSYS_MVX,lockerman2020livia,ahn2015scalable,nai2017graphpim,boroumand2018google,lazypim, top-pim, gao2016hrl, kim2018grim, drumond2017mondrian, RVU, NIM, PEI, gao2017tetris,Kim2016,gu2016leveraging, boroumand2019conda, hsieh2016transparent, cali2020genasm, NDC_ISPASS_2014,pattnaik2016scheduling,akin2015data,hsieh2016accelerating,lee2015bssync,boroumand2021mitigating,boroumand2021google,boroumand2022polynesia,boroumand2021polynesia,amiraliphd,besta2021sisa,fernandez2020natsa,singh2019napel,kwon202125,lee2021hardware,niu2022184qps,Sparse_MM_LiM,azarkhish2016logic,azarkhish2018neurostream,guo20143d,de2018design,akin2014hamlet,huang2020heterogeneous,dai2018graphh,liu2018processing,tsai:micro:2018:ams,gu2020ipim,DRAMA_CAL_2014,Asghari-Moghaddam_2016,huang2019active,kersey2017lightweight,li2019pims,kim2017grim,boroumand2017lazypim,zhuo2019graphq,zhang2018graphp,lim2017triple,smc_sim,HIVE,jang2019charon,IBM_ActiveCube,hadidi2017cairo,santos2018processing}\xspace}

\newcommand\pum{\cite{Chi2016, Shafiee2016, seshadri2017ambit, seshadri2019dram, li2017drisa, seshadri2013rowclone, seshadri2016processing, deng2018dracc, xin2020elp2im, song2018graphr, song2017pipelayer,gao2019computedram, eckert2018neural, aga2017compute,dualitycache,besta2021sisa,seshadri2016buddy,seshadri.bookchapter17,seshadri2018rowclone,seshadri2015fast,li2016pinatubo,ferreira2021pluto,ferreira2022pluto,imani2019floatpim,he2020sparse,flashcosmos,truong2022adapting,truong2021racer,olgun2021quactrng,kim2019d,kim2018dram,bostanci2022dr,olgun2022pidram,ali2019memory,angizi2019graphide,li2018scope,subramaniyan2017parallel,zha2020hyper,fujiki2018memory,orosa2021codic,sharad2013ultra,rezaei2020nom}\xspace}

\newcommand\drampum{\cite{angizi2019graphide,besta2021sisa,bostanci2022dr,deng2018dracc,ferreira2021pluto,ferreira2022pluto,gao2019computedram,li2017drisa,li2018scope,olgun2021quactrng,olgun2022pidram,seshadri.bookchapter17, seshadri2013rowclone,seshadri2015fast,seshadri2016buddy, seshadri2016processing, seshadri2017ambit,seshadri2018rowclone, seshadri2019dram, xin2020elp2im,missingnot,mimdramextended}\xspace}

\newcommand\drambackground{\cite{hassan2019crow,ghose.sigmetrics20, ghose2018your, kim2016ramulator, seshadri2019dram, kim2012case, zhang2014half, hassan2016chargecache, Tiered-Latency_LEE, seshadri2017ambit, chang2017understanding, chang2017understandingphd,
chang.sigmetrics2016, chang2014improving, chang2016low, lee2015adaptive, lee2016reducing, lee2016reducingthesis, lee2015decoupled, liu2013experimental, liu2012raidr, seshadri2013rowclone, seshadri2015gather, ipek2008self, lee2016simultaneous, Dennard68field,keeth2007dram,mineshphd,hasanphd,o2021energy}\xspace}

\newcommandx{\unsure}[2][1=]{\todo[linecolor=red,backgroundcolor=red!25,bordercolor=red,#1, size=\tiny]{#2}}
\newcommandx{\change}[2][1=]{\todo[linecolor=blue,backgroundcolor=blue!25,bordercolor=blue,#1,size=\tiny]{\textbf{#2}}}
\newcommandx{\feedback}[2][1=]{\todo[linecolor=yellow,backgroundcolor=yellow!25,bordercolor=yellow,#1]{#2}}
\newcommandx{\improvement}[2][1=]{\todo[linecolor=Plum,backgroundcolor=Plum!25,bordercolor=Plum,#1]{#2}}
\newcommandx{\thiswillnotshow}[2][1=]{\todo[disable,#1]{#2}}
\newcommandx{\completedRevision}[2][1=]{\todo[disable,backgroundcolor=red,#1]{#2}}
\newcommandx{\dataSource}[2][1=]{\todo[disable,backgroundcolor=red,#1]{#2}}
\newcommandx{\info}[2][1=]{\todo[linecolor=ddgreen,backgroundcolor=ddgreen!25,bordercolor=ddgreen,#1, size=\tiny]{#2}}

\newcommand{\boxbegin} {
	\begin{tcolorbox}[enhanced, frame hidden, colback=gray!50, breakable]
}

\newcommand{\boxend} {
	\end{tcolorbox}
}

\definecolor{lightblue}{rgb}{0.980, 0.956, 0.623}
\sethlcolor{lightblue}

\newcommand{\yboxbegin} {
	\begin{tcolorbox}[breakable, enhanced, frame hidden,
	enlarge top by=-0.25cm,
   enlarge bottom by=-0.1cm,
	colback=yellow!50]
}

\newcommand{\yboxend} {
	\end{tcolorbox}
}

\makeatletter
\patchcmd{\SOUL@ulunderline}{\dimen@}{\SOUL@dimen}{}{}
\patchcmd{\SOUL@ulunderline}{\dimen@}{\SOUL@dimen}{}{}
\patchcmd{\SOUL@ulunderline}{\dimen@}{\SOUL@dimen}{}{}
\newdimen\SOUL@dimen
\makeatother

\ifcameraready
\else
    \fancypagestyle{firstpage}
    {
        \fancyhead{}
        \fancyhead[C]{\textcolor{red}{CONFIDENTIAL DRAFT -- DO NOT DISTRIBUTE -- TO APPEAR IN HPCA'24} \\ \textcolor{MidnightBlue}{\emph{Version \versionnum~---~\today, \ampmtime}}  \hrule }
    }
\fi

\makeatletter
\def\bstctlcite{\@ifnextchar[{\@bstctlcite}{\@bstctlcite[@auxout]}}
\def\@bstctlcite[#1]#2{\@bsphack
 \@for\@citeb:=#2\do{%
   \edef\@citeb{\expandafter\@firstofone\@citeb}%
   \if@filesw\immediate\write\csname #1\endcsname{\string\citation{\@citeb}}\fi}%
 \@esphack}
\makeatother

\begin{document}
\bstctlcite{IEEEexample:BSTcontrol}

\title{\scalebox{0.8}{\prop: \omcriv{Enabling} High-Performance Processing-Using-DRAM with}\vspace{-5pt}
\scalebox{0.8}{Dynamic Bit-Precision, Adaptive Data Representation, and Flexible Arithmetic
}}

\author{
Geraldo F. Oliveira$^\dagger$~\quad
Mayank Kabra$^\dagger$~\quad
Yuxin Guo$^\ddagger$~\quad
Kangqi Chen$^\dagger$\\
A. Giray Ya\u{g}l{\i}k\c{c}{\i}$^\dagger$~\quad
Melina Soysal$^\dagger$~\quad
Mohammad Sadrosadati$^\dagger$\\
Joaquin O. Bueno$^\star$~\quad
Saugata Ghose$^\nabla$~\quad
Juan Gómez-Luna\textsuperscript{\S}~\quad 
Onur Mutlu$^\dagger$\vspace{10pt}
\\
$^\dagger$~\emph{ETH Zürich} \qquad 
$^\ddagger$~\emph{Cambridge University} \qquad 
$^\star$~\emph{Universidad de Córdoba} \\
$^\nabla$~\emph{Univ. of Illinois Urbana-Champaign} \quad
\textsuperscript{\S}~\it{NVIDIA Research}
\vspace{5pt}
}

\ifcameraready
 \thispagestyle{plain}
\else
  \thispagestyle{firstpage}
\fi
\pagestyle{plain}

\maketitle

\glsresetall

\begin{abstract}
%
\Gls{PuD} is \om{a paradigm where the analog operational properties of DRAM are used to perform \sgi{bulk} logic operations.}
\revdel{PIM trend that proposes to compute using the analog operation of DRAM cells.}%
\sgdel{Due to DRAM's \om{large} internal parallelism, }%
\sgi{While} \gls{PuD} promises high throughput at low energy and area cost,
\sgi{we uncover three limitations of existing \gls{PuD} approaches that lead to significant inefficiencies:} 
\li~static data representation\om{, i.e., \omcri{two}'s complement \sgdel{data representation}%
 with fixed bit-precision, leading to \emph{unnecessary computation} over useless \asplosrev{\hl{(i.e., inconsequential)}} data}; 
\lii~support for \emph{only} throughput-oriented execution\om{, where the high latency of individual \gls{PuD} operations can \emph{only} be hidden in the presence of bulk data-level parallelism;} and 
\liii~high latency for high-precision \om{(e.g., 32-bit) operations}. 
\revdel{\sgi{To address these issues, we propose \prop,}
\sgdel{To address these issues, we propose \prop, an adaptive \emph{data-representation} and \emph{operation-implementation} framework.}%
\om{which builds on two \emph{key ideas}. 
First, \prop \emph{parallelizes} the execution of independent primitives in a \gls{PuD} operation by leveraging DRAM's internal parallelism. 
Second, \prop \emph{reduces} the bit-precision for \gls{PuD} operations by leveraging \emph{narrow values}.}
\revdel{\sgi{We design an adaptive framework for \prop that uses hardware support at runtime, \emph{transparent} from the user, to select}
\li~the most efficient data format (e.g., 2's complement, redundant binary), 
\lii~the exact \sgi{bit-precision for a workload}, and
\sgdel{data precision that a workload requires (e.g., non-power-of-two integers), and}%
\liii~the fastest algorithm (e.g., bit-serial, bit-parallel) for latency- or throughput-oriented execution. }}

\gfcr{To address these issues, we propose \prop, the first hardware framework that addresses the high execution latency of bulk bitwise \gls{PuD} operations by implementing a data-aware runtime engine for \gls{PuD}. 
\prop~reduces the latency of \gls{PuD} operations in three different ways:
\li~\prop~\emph{dynamically} reduces the bit-precision (and \omcri{thus} the latency and energy consumption) of \gls{PuD} operations by exploiting narrow values (i.e., values with many leading zeros or ones);
\lii~\prop~\emph{concurrently executes} independent in-DRAM primitives \omcri{belonging} to a \emph{single} \gls{PuD} operation across \omcri{\emph{multiple}} DRAM arrays;  
\liii~\prop \emph{chooses and uses} the most appropriate data representation and arithmetic algorithm implementation for a given \gls{PuD} instruction \emph{transparently} to the programmer.}

%
\gf{We compare \prop to different state-of-the-art computing platforms (CPU, GPU, and the SIMDRAM \gls{PuD} architecture) \gfisca{for \om {twelve} real-world applications.} 
\omcri{Even when using only} a \omcri{\emph{single}} DRAM bank, 
\gfisca{\prop provides 
\li~17$\times$, 7.3$\times$, and 10.2$\times$ \omcri{higher} performance per mm$^2$; and
\lii~90.3$\times$, 21$\times$, and 8.1$\times$ \om{lower energy consumption than} CPU, GPU, and SIMDRAM, respectively, on average across \om{twelve} real-world applications.
\prop incurs low area cost on top of a DRAM chip (1.6\%) and CPU die (0.03\%).}}
\omcriv{We open-source \prop at \url{https://github.com/CMU-SAFARI/Proteus}.}

\end{abstract}
\glsresetall

\section{Introduction}
\label{sec:introduction}

\gf{\Gls{PIM}~\pimdef 
\sgdel{is a promising paradigm that}%
aims to alleviate the ever-growing cost of moving data\revdel{ back-and-forth} \juan{between} computing \omcri{units} (e.g., CPU, GPU) and memory \omcri{units} (e.g., DRAM). In \gls{PIM} architectures, computation is done by adding logic \emph{near} memory arrays, i.e., \gls{PnM}~\pnm, or by \emph{using} the analog \omcri{operational} properties of the memory arrays, i.e., \gls{PuM}~\pum). 
Prior works~\drampum show the \omcri{potential and}
\juan{feasibility of} \gls{PuD}, \omcri{by using} DRAM \omcri{circuitry} to implement \gfasplos{in-DRAM row copy~\omcri{\cite{seshadri2013rowclone,gao2019computedram,olgun2022pidram,yuksel2024simultaneous}}, Boolean~\omcri{\cite{li2017drisa,seshadri2017ambit,xin2020elp2im,seshadri2015fast}}, and arithmetic~\omcri{\cite{hajinazarsimdram,deng2018dracc,angizi2019graphide,li2018scope,peng2023chopper,mimdramextended,li2017drisa,zhou2022transpim,ali2019memory,park2024attacc,ferreira2022pluto,ferreira2021pluto,deng2019lacc,angizi2019redram,shin2024processing}} operations.}\revdel{ a variety of \gls{PuM} operations\revdel{, including bitwise Boolean~\cite{seshadri2017ambit, gao2019computedram, xin2020elp2im, besta2021sisa, li2017drisa} and arithmetic operations~\cite{deng2018dracc, gao2019computedram,li2017drisa,angizi2019graphide, hajinazarsimdram,li2018scope}}.}} 
\gf{\revdel{\gls{PuD} architectures leverage the key observations that 
\li~DRAM can perform Boolean \sgi{primitives} (e.g., NOR, MAJ) by simultaneously accessing (i.e., activating) multiple \sgdel{(e.g., three)}%
DRAM rows, and
\lii~complex arithmetic operations (e.g., addition) can be built by composing sequences of Boolean \sgi{primitives}.} 
\revdel{\sgi{\gls{PuD} frameworks~\cite{hajinazarsimdram, peng2023chopper} automate the implementation of complex arithmetic} operations using in-DRAM Boolean operations.} 
\sgi{\gfmicro{\gls{PuD} systems often} employ a \emph{bulk bit-serial} execution model\omcri{~\cite{seshadri2019dram}}, where each Boolean primitive operates across entire DRAM rows, with each row containing one bit from many input operands.
The predefined sequence of DRAM commands that implements an operation are stored in a \emph{\uprog}~\gfcri{\cite{hajinazarsimdram}}.}
\sgdel{Such frameworks employ a  \emph{bulk bit-serial} execution model. In this execution model, a \gls{PuD} operation is performed by \emph{sequentially} activating different DRAM rows, each of which stores one bit from the input operand, one row (i.e., a bit from the input operand) per time, following a predefined sequence of DRAM commands (called a \emph{\uprog}~\cite{hajinazarsimdram}). 
Thus, the number of bits used to represent operands (i.e., the \emph{bit-precision}) directly impacts the latency of a \gls{PuD} operation.
\sgi{Existing \gls{PuD} architectures employ a fixed-width bit representation to simplify execution.}}}

\sgi{\revdel{Unfortunately, while prior works have demonstrated the benefits of \gls{PuD} for several applications~\cite{seshadri2017ambit, xin2020elp2im, besta2021sisa,deng2018dracc, gao2019computedram,li2017drisa,angizi2019graphide, hajinazarsimdram,li2018scope},}
\gfcri{W}e uncover three \omcrii{shortcomings} that significantly \omcrii{limit} the performance and efficiency of \gls{PuD} architectures.
\revdel{First, they employ a \emph{static bit-precision} \gfmicro{for operands} (i.e., the number of bits used to represent an operand), which is wider than the maximum data values being computed on.
This results in significant wasted latency and energy to perform calculations on inconsequential bits (e.g., \asplosrev{\hl{leading zeros in case of positive numbers or leading ones in case of negative numbers}}).
Second, they aim to \emph{maximize execution throughput} for large data sets, \gfasplos{to} overcome long bit-serial latencies and take advantage of the large amount of in-DRAM parallelism.
Whenever these architectures execute on smaller data sets, much of this parallelism \asplosrev{\hl{is not fully utilized}},
\emph{even though there are opportunities to better exploit this parallelism to improve latency and efficiency.}
Third, they suffer from \emph{high latencies for high-precision computation} due to their bit serialization.
\revdel{Forcing bit-serial computation instead of leveraging alternate execution models can significantly harm program efficiency.}
Ultimately, these limitations exist because there is no one-size-fits-all solution for data representation and execution model across a wide range of applications.}}
\gfcr{First, they employ a \textbf{rigid and static data representation}, which is \emph{inefficient} for bit-serial execution.
Existing \gls{PuD} engines typically employ a \omcri{\emph{fixed}} bit-precision, statically\omcri{-}defined \omcri{(commonly two's complement)} data representation for \omcri{\emph{all}} \gls{PuD} operations. 
This rigid and static data format introduces inefficiencies in a bit-serial execution model, where bits of a data word are \emph{individually} and \emph{sequentially} processed. 
Since many applications store data in data representation formats that exceed the necessary precision~\omcri{\cite{pekhimenko2012base,alameldeen2004adaptive,islam2010characterization,ergin2006exploiting,brooks1999dynamically,ergin2004register,budiu2000bitvalue,wilson1999case,gena-thesis,pekhimenko2013linearly}} (e.g., 8-bit values stored in a 32-bit integer), a significant fraction of \gls{PuD} computation is wasted on \omcri{processing} inconsequential bits, such as leading zeros or \omcri{ones (e.g., sign-\omcrii{extension} bits)}, causing significant latency and energy overhead.
Second, \gls{PuD} architectures \omcri{provide} \emph{only} a \textbf{throughput-oriented execution \omcri{model} with limited latency tolerance}.
\gls{PuD} operations are composed of \omcri{bitwise (bit-serial)} in-DRAM primitives (e.g., in-DRAM majority \omcri{and \texttt{NOT}~\cite{seshadri2017ambit,seshadri2015fast}}), making individual \gls{PuD} operations inherently slow \omcri{due to the need to operate on each bit \omcrii{serially} to perform an operation with a data width larger than one}. 
To compensate for this latency, \gls{PuD} architectures adopt a \emph{throughput-oriented execution model} that distributes large amounts of data across multiple DRAM subarrays and DRAM banks, enabling massively parallel execution \omcri{on many data elements}.
However, this approach is effective \omcri{\emph{only}}  when sufficient data-level parallelism is available to amortize the high latency of \omcrii{bit-serial} \omcri{execution of an} individual \gls{PuD} \omcri{primitive}. 
In scenarios where data-level parallelism is limited, this throughput-oriented execution model fails to hide the latency of individual in-DRAM primitives, potentially leading to performance degradation~\omcri{\cite{mimdramextended}}. 
Third, bit-serial \gls{PuD} architectures face \textbf{scalability challenges for high-precision operations}. 
\gls{PuD} systems suffer from increased latency as the target bit-precision grows~\omcri{\cite{hajinazarsimdram}}. 
Due to their bit-serial nature, the \omcrii{latencies} of arithmetic \gls{PuD} operations \omcrii{scale} linearly~\cite{hajinazarsimdram} \omcri{(e.g., for addition/subtraction) or quadratically~\cite{hajinazarsimdram} (e.g., for multiplication/division)} with the target bit-precision.}

\sgdel{\gf{Even though prior works have demonstrated the performance and energy benefits of employing \gls{PuD} for several applications, they suffer from \emph{three} main limitations caused by naively using a bit-serial execution model, which in turn,  hinder their performance. 
First, they \emph{undermine} the full benefits of the underlying bit-serial execution model by using a \emph{static data representation} for \gls{PuD} computation. 
This is because prior works utilize conventional data formats (e.g., two's complement) and operands' bit-precision (e.g., 32-bit integers) for \gls{PuD} computation even when computation could be performed with fewer bits, leading to subpar performance.
Second, they \emph{only} support a \emph{throughput-oriented execution}, which can hurt the overall performance of latency-sensitive operations. \gls{PuD} architecture often favors throughput-oriented execution because the bulk parallelism inside a DRAM array can 
\juan{amortize} the latency of consecutive DRAM activations in a \uprog. 
\revdel{To break even in terms of performance (compared to a processor-centric execution), the \gls{PuD} operation needs to operate over large input sets to fully utilize the bulk parallelism the underlying \gls{PuD} architecture offers.} When executing \gls{PuD} operations with smaller-than-ideal input set sizes \juan{(e.g., less than the DRAM row size)}, the \gls{PuD} architecture can experience sub-optimal performance or even performance loss compared to their processor-centric counterparts.    
Third, they suffer from \emph{high latency} when executing \emph{high-precision computation}. In a bit-serial execution model, the computation latency scales with bit-precision used for operands. This scaling can \emph{significantly} degrade performance when using a large bit-precision (e.g., 64-bits) for quadratically-scaling bit-serial \gls{PuD} operations \gfisca{(e.g., multiplication, division)}.}}

\gf{Our \emph{goal} in this work is to overcome the three limitations of \gls{PuD} architectures \omcrii{that stem from} \omcri{the naive use of} a bit-serial execution model. To this end, we propose \prop,\footnote{\gfasplos{\prop is a shape-shifting, prophetic sea god from Greek mythology~\omcri{\cite{homer_odyssey}}, known for his ability to elude capture by changing forms.} \omcri{Our} \prop \emph{changes} the bit-precision of \gls{PuD} operations to improve performance.} an efficient \gfcri{data-aware runtime framework that \emph{dynamically} adjusts the bit-precision and, based on that, \emph{chooses} and \emph{uses} the most appropriate data representation and arithmetic algorithm implementation for a given \gls{PuD} operation}.
\gfcr{\prop builds on three \emph{key ideas}. 
To solve the \textbf{first limitation} (i.e., rigid and static data representation), \prop \emph{reduces} the bit-precision for \gls{PuD} operations by leveraging \emph{narrow values} (i.e., values with many leading zeros \omcri{or ones}).
As several works observe~\omcri{\cite{pekhimenko2012base,alameldeen2004adaptive,islam2010characterization,ergin2006exploiting,brooks1999dynamically,ergin2004register,budiu2000bitvalue,wilson1999case,gena-thesis,pekhimenko2013linearly}}, programmers often over-provision the bit-precision used to store operands, using large data types (e.g., a 32-bit or 64-bit integer) to store small (i.e., narrow\omcri{, e.g., 4-bit, 8-bit}) values. 
Based on this observation, \prop \omcri{exploits dynamic} narrow values to reduce the bit-precision of a \gls{PuD} operation to that of the best-fitting number of bits\omcri{, thereby avoiding} costly in-DRAM operations \omcri{on inconsequential} bits, which improves overall performance and energy efficiency.}

\gfcr{To solve the \textbf{second limitation} (i.e., throughput-oriented execution with limited latency tolerance), \prop \emph{parallelizes} the execution of \emph{independent} in-DRAM primitives in a \gls{PuD} operation by leveraging DRAM's internal organization combined with \emph{bit-level parallelism}.
We make the \emph{key observation} that many in-DRAM primitives that compose a \gls{PuD} operation (e.g., an in-DRAM addition) can be executed \emph{concurrently} across different bits of a data word.
For example, executing an $n$-bit in-DRAM addition (i.e., \gfcri{$\{{A_{n-1}, \dots, A_0}\} + \{B_{n-1}, \dots, B_0\}$}) in a bit-serial manner requires \omcri{serially} performing three majority-of-three (\texttt{MAJ3}) operations per bit $i$ to compute the $sum$ and propagate the carry to bit $i{+}1$. 
However, only one of these \omcri{operations} (i.e., the carry propagation from bit $i$ to bit $i+1$) \omcri{truly requires serialization}, while the other two \texttt{MAJ3} operations can be \emph{concurrently executed} \gfcri{\omcrii{\emph{across}} the $n$ bit positions of a data word}.
To exploit this observation, \prop scatters the $n$ bits of a data word across \omcri{\emph{multiple}} DRAM subarrays \gfcri{(i.e., $subarray_0 \leftarrow  \{A_0,B_0\}, \dots$, $subarray_{n-1} \leftarrow  \{A_{n-1},B_{n-1}\}$}) and employs \gls{SLP}~\cite{kim2012case} to enable each subarray $i$  to \emph{concurrently} execute the in-DRAM primitive associated with bit $i$, \omcri{thereby} hiding the high latency of individual in-DRAM primitives in a \gls{PuD} operation over the many bits of the target data word.}
\gfcrii{To propagate intermediate data (e.g., carry bits) across DRAM subarrays, \prop leverages LISA~\cite{chang2016low}, a low-cost \omcriii{DRAM design} that enables fast inter-subarray data movement at DRAM row granularity.}

\gfcr{To solve the \textbf{third limitation} (i.e., scalability challenges for high-precision operations), \prop \omcri{exploits} an alternative data representation for high-precision computation. 
Concretely, we \omcri{use} the \omcrii{\emph{\gls{RBR}}}~\cite{guest1980truth,phatak1994hybrid,lapointe1993systematic, olivares2006sad, olivares2004minimum} (where multiple-digit combinations represent the same value), for high-precision \gfcri{(e.g., 32-bit or 64-bit)} \omcri{\gls{PuD}} \omcri{computations}. 
\gls{PuD} execution can
take advantage of two properties of \gls{RBR}-based arithmetic: 
\li~operations no longer need to propagate carry bits through the full width of the data (e.g., \gls{RBR}-based addition limits carry propagation to at most two \omcri{digits}~\cite{brown2002using}), and 
\omcri{\lii}~operation latency is \emph{independent} of the bit-precision.
}

\revdel{The \emph{key ideas} behind \prop are: 
\li~fully leverage the parallelism \sgi{within} a DRAM bank to accelerate the execution of various bit-serial and bit-parallel arithmetic operations, and 
\lii~decide \sgi{at runtime} the best-performing bit-precision, data representation format, and algorithmic implementation of \sgi{each \gls{PuD}} operation. \prop is composed of three main components: \uproglib, \dynengine, and \uprogunit. \revdel{With these three components, \prop can adapt the data representation and \gls{PuD} operation implementation depending on the target bit-precision.}}}

\gfcr{Based on these \omcri{three key ideas}, we design \prop as a three-component hardware \gfcri{runtime} framework for high-performance \gls{PuD} computation that \emph{transparently} (from the user/programmer) selects\gfcri{, for a given \gls{PuD} operation, the} 
\li~bit-precision,
\lii~fastest arithmetic algorithm for latency- or throughput-oriented \gls{PuD} execution, and} 
\liii~most efficient data format (e.g., two's complement \omcri{or} redundant binary~\omcri{\cite{guest1980truth,phatak1994hybrid,lapointe1993systematic, olivares2006sad, olivares2004minimum}}).
\gf{First, we build a \uproglib, consisting of hand-tuned \gfcri{algorithmic} implementations of key arithmetic operations \gfcri{that take into account \gls{SLP}~\cite{kim2012case} to implement \gls{PuD} operations} using  
\li~\omcrii{both} bit-serial and bit-parallel algorithms and 
\lii~\omcrii{both} two's complement and \gls{RBR} data representation formats. 
\gfcri{The} \uproglib contains a collection of possible implementations of \gls{PuD} operations, each of which with different \gfcri{predetermined latency and energy requirements that} depend on \gfcri{a given} bit-precision. \gfcri{The latency and energy each \uprog consumes is stored within an easily-accessible \gfcriii{\prelut} alongside the \uprog in the \uproglib.}} 

\gf{Second, we devise a new \dynengine to identify the appropriate initial bit-precision for a given \gls{PuD} \gfcri{operation}. We implement the  \dynengine by augmenting prior works' \emph{Data Transposition Unit}\juan{~\cite{hajinazarsimdram,mimdramextended}}. 
Before \gls{PuD} execution, the \emph{Data Transposition Unit} captures and transposes (from the standard horizontal data layout to the \gls{PuD} vertical data layout) cache lines that are about to be evicted from the \gls{LLC} to DRAM and belong to a \gls{PuD} memory object, \gfcri{which is \emph{previously-identified} \gfcri{based on its memory address range}}. 
During this process, our  \dynengine \emph{\gfmicro{scans}} the content of the evicted cache line to identify the largest value belonging to \gfasplos{a} \gls{PuD} memory object.
Third, when a \gls{PuD} \gfcri{operation} is issued, the \uprogunit probes the \dynengine to identify the most suitable bit-precision for the \gls{PuD} \gfcri{operation} and, \gfcri{based on the \prelut within the \uprogunit}, selects the best performing \uprog from the \uproglib. }

\paratitle{\omcri{Key Results}} \gf{We compare \prop to different state-of-the-art computing platforms (\omcri{state-of-the-art} CPU, GPU, and SIMDRAM~\cite{hajinazarsimdram}). 
We comprehensively evaluate \prop'
performance for \gfisca{twelve} real-world applications~\omcrii{\cite{pouchet2012polybench,spec2017,yoo_iiswc2009,che_iiswc2009}}. 
\omcri{Even when using only a \emph{single} DRAM bank,} 
\gfisca{\prop provides 
\li~17$\times$, 7.3$\times$, and 10.2$\times$ \omcri{higher} performance per mm$^2$; and
\lii~90.3$\times$, 21$\times$, and 8.1$\times$ \om{lower energy consumption than} CPU, GPU, and SIMDRAM, respectively, on average across all \om{twelve} real-world applications.}
\prop incurs low area cost on top of a DRAM chip (1.6\%) and CPU die (0.03\%).}

\gf{We make the following \omcri{major} contributions:}
\begin{itemize}
[noitemsep,topsep=0pt,parsep=0pt,partopsep=0pt,labelindent=0pt,itemindent=0pt,leftmargin=*]
\item \gfcri{We identify three \omcrii{major shortcomings} of existing bit-serial \gls{PuD} architectures that significantly \omcrii{limit} performance and energy efficiency: \li~rigid and static data representation, 
\lii~throughput-oriented execution model with limited latency tolerance, and \liii~scalability challenges for high-precision operations.}

\item \gfcri{We propose \prop, a \gfcrii{three-component} data-aware hardware runtime framework  \gfcrii{for high-performance \gls{PuD} computation} that \gfcrii{\emph{transparently} (from the programmer), for a given \gls{PuD} operation,} \emph{dynamically} \omcrii{adjusts} the bit-precision, and based on that, \emph{chooses} and \emph{uses} the most appropriate data representation \gfcrii{(e.g., two's complement or \gls{RBR})} and \omcriii{the most appropriate} arithmetic algorithm implementation \gfcrii{for latency- or throughput-oriented \gls{PuD} execution}.} 

\item \gf{We extensively evaluate \prop \agy{for} twelve real-world applications\agy{, showing} that \prop 
outperforms \gfcr{a} state-of-the-art \gls{PuD} framework \omcri{ \omcrii{(SIMDRAM)}, CPU, and GPU} while incurring low area \agy{cost} to the system.}

\item \gfcriv{We open-source \prop at \url{https://github.com/CMU-SAFARI/Proteus}.}
\end{itemize}

\section{Background}
\label{sec:background}

\revdel{We first briefly explain the architecture of a typical DRAM chip. Next, we describe prior DRAM enhancements and \gls{PuD} works that \prop builds on top of.}

\subsection{\gf{DRAM Organization \& Operation}} 
\label{sec:background:organization}

\begin{figure}[ht]
    \centering
    \includegraphics[width=\linewidth]{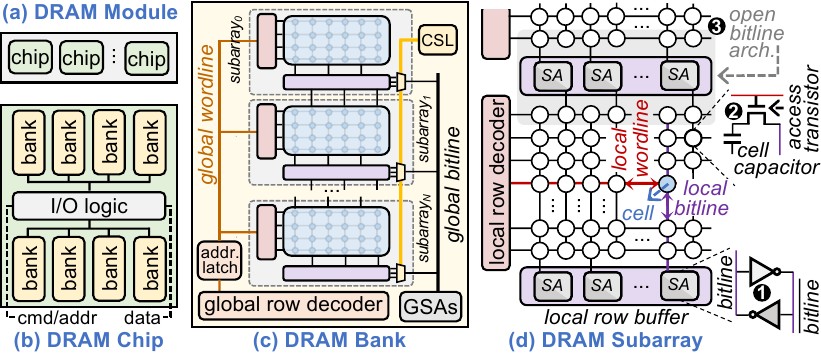}
    \caption{\omcrii{DRAM organization.}}
    \label{fig_subarray_dram}
\end{figure}

\paratitle{\gfisca{DRAM Organization}} 
\sgi{Fig.~\ref{fig_subarray_dram} shows the hierarchy of a DRAM system.}
\gf{%
\sgdel{A DRAM system comprises a hierarchy of components, as Fig.~\ref{fig_subarray_dram} illustrates.}
A \emph{DRAM module} (Fig.~\ref{fig_subarray_dram}a) has several (e.g., 8--16) DRAM chips. 
A \emph{DRAM chip} (Fig.~\ref{fig_subarray_dram}b) has multiple banks (e.g., 8--16). 
A \emph{DRAM bank} (Fig.~\ref{fig_subarray_dram}c) has 
\li~multiple (e.g., 64--128) 2D arrays of DRAM cells known as \emph{DRAM subarrays}~\omcri{\drambackground};
\lii~a \emph{global row decoder} and a \emph{global address latch} that select a row of cells in a subarray through \emph{global wordlines};
\liii~\emph{column select logic} (CSL) that selects portions (e.g., 64-bit) of the row; and
\liv~\gfcr{a set of }\sgdel{\emph{global sense amplifier} \sg{(i.e., a \emph{global row buffer}, GRB)}}%
\emph{global \gfcr{sense amplifiers}} \gfcr{(GSAs)~\omcri{\cite{lee2017design,lee2016simultaneous,seshadri2016processing,chang2016low,seshadri2019dram,lee2016reducingthesis,vivekphd,chang2017understandingphd}}, also sometimes called \omcrii{the} global row buffer~\omcri{\cite{mimdramextended,wang2020figaro,ferreira2022pluto,kim2012case,olgun2024sectored,jeremiephd,kim2018solar}},} that transfers the selected fraction of the data from the row through \emph{global bitlines}.}
\sgi{Each subarray (Fig.~\ref{fig_subarray_dram}d) contains 
\li~multiple rows (e.g., 512--1024) and columns (e.g., 2--8~kB~\cite{kim2018solar, lee2017design, kim2002adaptive}\gfisca{)} of DRAM cells,
\lii~a \emph{local row decoder} that activates a \emph{local wordline}, and
\liii~a \emph{local row buffer} containing a row of \emph{sense amplifiers} (SAs; \circled{1} in Fig.~\ref{fig_subarray_dram}d) to latch data from an activated row.
A DRAM cell (\circled{2}) consists of an access capacitor, which connects a transistor that stores the data value with a \emph{local bitline} shared by all cells in the same column.
Modern DRAM employs an \emph{open bitline architecture}~\cite{lim20121,takahashi2001multigigabit}, fitting only enough SAs in one local row buffer to latch half a row of cells. To latch the entire row, a subarray \asplosrev{\hl{connects}} to \emph{two} local row buffers, one above the cell array and one below (\circled{3}).}

\sgdel{\gf{A \emph{DRAM subarray} (Fig.~\ref{fig_subarray_dram}d) organizes DRAM cells into multiple \emph{rows} (e.g., 512--1024) and multiple \emph{columns} (e.g., 2--8~kB~\cite{kim2018solar, lee2017design, kim2002adaptive}\gfisca{)} and contains
\li~a \emph{local row decoder} that drives the \emph{local wordlines} to the appropriate voltage levels to activate a row; and  
\lii~a row of sense amplifiers (also called a \emph{local row buffer}) that senses and latches data from the activated row.
A \emph{sense amplifier} \sg{(SA)} comprises of two back-to-back inverters (\circled{1} in Fig.~\ref{fig_subarray_dram}d).
A \emph{DRAM cell} (\circled{2}) consists of an \emph{access transistor} and a \emph{storage capacitor}. 
The source nodes of the access transistors of all the DRAM cells in the same column connect the cells' storage capacitors to the same \emph{local bitline}. The gate nodes of the access transistors of all the DRAM cells in the same row connect the cells' access transistors to the same \emph{local wordline}. 
To achieve high density, modern DRAM designs employ an \emph{open bitline architecture}~\cite{lim20121,takahashi2001multigigabit}, fitting only enough sense amplifiers in a local row buffer to sense half a row of cells. To sense the entire row of cells, each subarray has local bitlines connecting to two rows of local sense amplifiers -- one above and one below the cell array (\circled{3}).}}

\paratitle{\gf{DRAM Operation}} 
\sgi{The memory controller issues three commands to service a DRAM request.
\revdel{Initially, the local bitlines are set at a reference voltage.}
The first command, \texttt{ACTIVATE} (\texttt{ACT}),\revdel{ selects a specified DRAM row.
This} connects \omcrii{each} DRAM \omcrii{cell} in \gfmicro{a} row to its local bitline,
and the cell's transistor shares its charge with the bitline to
shift the bitline voltage higher (or lower) if the cell stores a `1' (`0').
The local row buffer amplifies the shifts to CMOS-readable values (simultaneously restoring charge to the DRAM cell).
The latency from the start of activation until charge restoration is called $t_{RAS}$.
The second command, \texttt{READ} (\texttt{RD}), returns a cache line of data from the \gfmicro{local row buffer}.
The third command, \texttt{PRECHARGE} (\texttt{PRE}), disconnects DRAM cells from the bitlines, and returns the bitlines to their reference voltage.
The precharge latency is called $t_{RP}$.}
\sgdel{\gf{Three major steps are involved in serving a main memory request. First, to select a DRAM row, the memory controller issues an \texttt{ACTIVATION} (\texttt{ACT}) command with the row address. On receiving this command, DRAM transfers all the data in the row to the corresponding local row buffers. \sgdel{(i.e., the row buffer at the top of the subarray and the one at the bottom).}
The two-terminal sense amplifier in the local row buffers senses the voltage difference between the local bitline and a reference voltage and amplifies it
to a CMOS-readable value \gfisca{until the cell charge is restored}. 
\gfisca{The latency from the start of row activation until the completion of the DRAM cell's charge restoration is called \emph{charge restoration latency} ($t_{RAS}$).}
Second, to access a cache line from the activated row, the memory controller issues a \texttt{READ} (\texttt{RD}) command with the column address of the request. 
Third, to enable the access of another DRAM row in the same bank, the memory controller issues a  \texttt{PRECHARGE} (\texttt{PRE}) command. 
This command disconnects the local bitline and restores the local bitline voltage to its quiescent state. \gfisca{The latency between issuing a \texttt{PRE} and when the DRAM bank is ready for a new row activation is called \emph{precharge latency} ($t_{RP}$)}.}}

\revdel{\paratitle{\gf{Bank and Subarray-Level Parallelism}} \gf{Modern DRAM systems introduce parallelism at varying levels of the DRAM hierarchy. In particular, a single DRAM chip can leverage \emph{\gls{BLP}} and \emph{\gls{SLP}}. First, \gls{BLP} is the ability to serve memory requests targeting different DRAM banks \emph{concurrently}. This is possible since each DRAM bank in a DRAM chip has its own dedicated global peripheral accessing circuitry\revdel{ (i.e., a global row decoder, global sense amplifier, and CSL)}, which can be used to drive different memory requests. \revdel{The memory controller is then responsible for orchestrating the utilization of DRAM resources shared across all DRAM banks in a DRAM module (i.e., I/O logic and memory channel).}
Second, \gls{SLP} is the ability to serve memory requests targeting different DRAM subarrays in a single DRAM bank \emph{concurrently}. Different than \gls{BLP}, implementing \gls{SLP} is \emph{not} trivial since memory requests targeting different subarrays in a single DRAM bank are \emph{serialized}. \revdel{The serialization happens because all subarrays in a single DRAM bank share the same global peripheral accessing circuitry.} 
To implement \gls{SLP}, SALP~\cite{kim2012case} introduces simple modifications to the global peripheral accessing circuitry of a DRAM bank that allows for concurrent access to 
\li~one subarray using \emph{SALP-1}, which overlaps the precharging of one subarray with the activation of another subarray; 
\lii~two subarrays using \emph{SALP-2}, which issues an \texttt{ACTIVATE} to another subarray before the \texttt{PRECHARGE} to the currently activated subarray; and 
\liii~multiple subarrays using MASA, which allows the memory controller to designate exactly one of the activated subarrays to drive the global bitlines during the next column command using a new DRAM \gfisca{command} called \texttt{SA\_SEL}.       }}

\subsection{\gf{Processing-\sgi{Using}-DRAM}} 
\label{sec:background:PUD}

\paratitle{\gf{In-DRAM Row Copy}} \gf{RowClone~\cite{seshadri2013rowclone}  enables copying a row~$A$ to a row~$B$ in the \emph{same} subarray by issuing two consecutive \texttt{ACT} commands to these two rows, followed by a \texttt{PRE} command. This command sequence is called \texttt{AAP}.
LISA~\cite{chang2016low} enables the execution of in-DRAM row copy operations across DRAM rows in \emph{different} subarrays of a DRAM chip by connecting local row buffers of \omcrii{neighboring} subarrays using isolation transistors. 
}

\paratitle{\gf{In-DRAM Bitwise Operations}} 
\gf{\sgi{Ambit~\omcrii{\cite{seshadri2017ambit,seshadri2015fast}} shows that a simultaneous \emph{{\gls{TRA}}} can perform \emph{in-DRAM} bitwise \omcrii{\texttt{MAJ3}/}\texttt{AND}/\texttt{OR} operations.}
\revdel{When activating three rows, three cells connected to each local bitline share charge simultaneously and contribute to the perturbation of the local bitline. Upon sensing the perturbation of the three simultaneously activated rows, the sense amplifier amplifies the local bitline voltage to $V_{DD}$ or 0 if at least two of the capacitors of the three DRAM cells are charged or discharged, respectively.}\revdel{ As such, a \gls{TRA} results in a Boolean majority operation ($MAJ$).}
\sgdel{Ambit implements \gls{TRA} by introducing a custom row decoder to perform a \gls{TRA} by simultaneously addressing three wordlines. To use this decoder, Ambit 
defines a new command called \texttt{AP} that issues a \gls{TRA} followed by a \texttt{PRE}.}%
\sgi{Ambit implements \gls{TRA} using a custom row decoder, and introduces a new command called \texttt{AP} that issues a \gls{TRA} followed by a \texttt{PRE}.
\omcrii{Ambit also provides a mechanism to perform bitwise \texttt{NOT} operations~\cite{seshadri2017ambit}.}
SIMDRAM~\cite{hajinazarsimdram}\revdel{ is a framework that} builds on top of Ambit to implement and expose high-level in-DRAM operations (\emph{{\uprog}s}).
A \uprog consists of a sequence of \texttt{AAP}s (row copies) and \texttt{AP}s that are generated offline, and exposed to the programmer as \omcrii{new} \emph{bbop} \omcrii{(bulk-bitwise operation)} instructions.
\revdel{During program execution, SIMDRAM
\li~decodes a \emph{bbop} instruction into its respective \uprog and
\lii~dispatches the \aaps in the \uprog to DRAM.}
To implement carry propagation, SIMDRAM employs a \emph{vertical} data layout, where all bits of a data word\revdel{ (e.g., a 32-bit integer)} are stored in a single DRAM column, and executes \omcrii{each \emph{bbop} instruction} bit-serially\revdel{(i.e., one row at a time)}.}
\asplosrev{\hl{Such \omcrii{an} execution model allows SIMDRAM to perform \emph{implicit bit-shift} operations via in-DRAM row copies, e.g., SIMDRAM performs a left-shift-by-one operation by copying the data in DRAM row $j$ to DRAM row $j+1$.}}
\gfcriii{Across this paper, we use the following terminology:
\li~a \emph{\gls{PuD} operation} refers to the target computation that DRAM executes (e.g., addition, row copy);
\lii~an \emph{in-DRAM}/\emph{\gls{PuD} primitive} refers to the sequence of \aaps in a \uprog; and
\liii~a \emph{\gls{PuD} instruction} refers to a \emph{bbop} instruction that the user/compiler uses as the software interface to trigger a \gls{PuD} operation.
}
\sgdel{SIMDRAM~\cite{hajinazarsimdram} builds on top of Ambit by proposing a three-step framework that translates an operation into its in-DRAM representation called \emph{\uprog}. A \uprog consists of a sequence of \texttt{AAP}s (row copies) and \texttt{AP}s (\gls{TRA}s) that implements an operation (e.g., addition) in DRAM. In SIMDRAM, \uprogs are generated offline, stored in DRAM for future use, and exposed to the programmer as \emph{bbop} instruction. 
To implement complex arithmetic operations that require carry propagation (e.g., addition and multiplication), SIMDRAM employees a \emph{vertical} data layout, where all bits of a data word (e.g., a 32-bit integer) are stored in a single DRAM column. In this way, SIMDRAM implements \asplosrev{\hl{a}} \emph{bit-serial} computation. 
}}

\section{Motivation}
\label{sec_motivation}

\gfcr{We discuss three \omcrii{major shortcomings} of prior \gls{PuD} architectures: \emph{static data representation}, \emph{support for only throughput-oriented execution}, and \emph{high latency for high-precision operands.}
}

\paratitle{\omcrii{\emph{Limitation 1}:}~\gf{Static Data Representation}} 
\sgi{\Gls{PuD} architectures naively \omcrii{utilize} conventional data formats (e.g., two's complement) and fixed operand bit-precision (e.g., 32-bit integers) to implement bit-serial computation. However, because bit-serial latency \omcrii{directly increases} with bit-precision, these architectures experience subpar performance since \asplosrev{\hl{\omcrii{an} application's data with small dynamic range (i.e., narrow values) are often stored in large data formats}}~\cite{pekhimenko2012base,alameldeen2004adaptive,islam2010characterization,ergin2006exploiting,brooks1999dynamically,ergin2004register,budiu2000bitvalue,wilson1999case} that waste most of the bit-precision.} \asplosrev{\hl{Note that data values \omcrii{often} become narrow dynamically at runtime.}}
\gf{\sgdel{Prior bit-serial \gls{PuD} architectures \emph{undermine} the full benefits of the underlying bit-serial execution model by using a \emph{static data representation} for \gls{PuD} computation. This is because prior works naively utilize conventional data formats (e.g., two's complement) and operands' bit-precision (e.g., 32-bit integers) for \gls{PuD} computation, which leads to subpar performance. 
In a bit-serial execution model, the number of operations required to perform a given computation is tight to the bit-precision of the computation. \gfisca{Thus, using fewer bits to represent operands results in lower latency. 
One way to minimize the bit-precision for bit-serial \gls{PuD} computation is to exploit \emph{narrow values}. As several works observe~\cite{pekhimenko2012base,alameldeen2004adaptive,islam2010characterization,ergin2006exploiting,brooks1999dynamically,ergin2004register,budiu2000bitvalue,wilson1999case}, programmers often over-provision the bit-\gfisca{precision} used to store operands, using large data types (e.g., a 32-bit \sg{or 64-bit} integer) to store small \gfisca{(i.e., narrow}) values \sg{(e.g., an up to 255-pixel value in an RGB image)}.}}%
Narrow values have been exploited in many scenarios, \sgi{e.g.,} cache compression~\gfcrii{\cite{pekhimenko2012base,alameldeen2004adaptive,wilson1999case,islam2010characterization,duan2014exploiting,molina2003non,kong2012exploiting,pekhimenko2015toggle,pekhimenko2016case,pekhimenko2015exploiting,gena-thesis}}, 
register files~\gfcrii{\cite{ergin2004register,wang2017gpu,hu2006register,wang2009exploiting,ergin2006exploiting,ozsoy2010dynamic,mittal2017design,ergin2006exploitingpatmos}}, 
logic synthesis \gfcrii{\& circuit optimizations}~\cite{canesche2022polynomial,canis2013software,pilato2013bambu,onur2009exploiting,osmanlioglu2009reducing}\gfcrii{, 
neural network quantization~\cite{jang2022encore, albericio2017bit},
error tolerance~\cite{ergin2006exploiting,karsli2012enhanced,ergin2008reducing}}.}

\begin{figure}[ht]
    \centering
    \includegraphics[width=\linewidth]{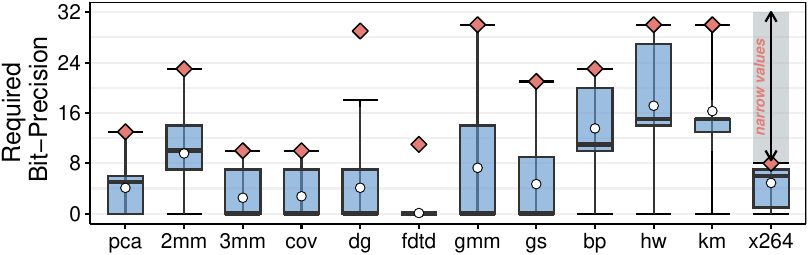}
    \caption{\revA{\gfisca{\sgi{\omcrii{Required b}it-precision distribution\revdel{ (i.e., minimum number of bits needed for an operand)} for \gfcrii{input/output data arrays of auto-vectorized arithmetic instructions in loops across} 12 applications.
    The box represents the 25th to 75th percentiles, with whiskers extending to the smallest/largest precision (with a diamond at the largest precision and a bubble at the mean precision).}
    \sgdel{Bit-precision distribution of twelve real-world applications.
    The y-axis indicates the \emph{required bit-precision} for a given \gls{PuD} operation, defined as the minimum number of bits required to represent input operands. Whiskers extend to the minimum and maximum data points on either side of the box. A bubble depicts average values. A diamond indicates the \emph{minimum} bit-precision that covers 100\% of the input operands.\prtagA{Fix.}}}}}
    \label{fig:narrow_values}
\end{figure}

\noindent \gf{\omcrii{\textbf{\emph{Opportunity 1:} Narrow Values for \gls{PuD} Computation.}}\revdel{In the context of \gls{PuD},} \gfmicro{N}arrow values \gfisca{can} be exploited to reduce the bit-precision of a \gls{PuD} operation to that of the best-fitting number of bits\omcrii{, thereby}, improving overall performance. 
\gfisca{We quantify the required bit-precision in \gls{PuD}-friendly real-world applications in Fig.~\ref{fig:narrow_values}. 
We define as \emph{required bit-precision} the minimum number of bits required to represent the input operands of the \gls{PuD} operation.\footnote{For example, if the input operand is an integer storing the value `\texttt{2}', the required bit-precision for such an input operand would be \sg{3} bits (\sg{two bits to represent the data \omcrii{value} and one bit to represent the sign}).} 
\revD{\changeD{D2}We collect the required bit-precision dynamically in three main steps: we 
\li~instrument \asplosrev{\hl{loops in applications that can be auto-vectorized using LLVM's loop auto-vectorization pass~\mbox{\cite{ lattner2008llvm,sarda2015llvm, lopes2014getting,writingpass}} (since prior work~\mbox{\cite{mimdramextended}} shows that such loops are \omcrii{well-suited} for \mbox{\gls{PuD}} execution)}} to output the data \omcrii{values} such \omcrii{loops use} \gfcrii{(i.e., we collect the data values of \emph{each} data array that is used as input/output of an auto-vectorized arithmetic instruction across the auto-vectorized loops in an application)}, 
\lii~execute the application to completion, \liii~post-process the output file containing the loop information data to calculate the required bit-precision.}}} 

{{We make two observations.
First, \gfisca{all our real-world applications display a \emph{significant} amount of narrow values. 
In such applications, the input bit-precision can be reduced from the native 32-bit to 20-bit (min. of 8-bit, max. of 30-bit)} \sg{on average across \emph{all} applications}. 
By doing so, the performance of the underlying \gls{PuD} architecture can improve by 1.6$\times$, \sg{in case the application utilizes linearly-scaling \gls{PuD} operations (such as addition~\omcrii{\cite{hajinazarsimdram}}),} or 2.6$\times$, \sg{in case the application utilizes quadratically-scaling \gls{PuD} operations (such as multiplication~\omcrii{\cite{hajinazarsimdram}})}. 
Second, the bit-precision significantly varies \omcrii{across data arrays within} a given application. This indicates the need for a mechanism that can \emph{dynamically} identify the target bit-precision for a given \gls{PuD} operation \gfmicro{(similar to prior works that leverage narrow values for tasks other than \gls{PuD}~\cite{pekhimenko2012base,alameldeen2004adaptive,islam2010characterization,ergin2006exploiting,brooks1999dynamically,ergin2004register,budiu2000bitvalue,wilson1999case,hu2006register,wang2017gpu,lipasti2004physical,loh2002exploiting})}.}}
\revD{\label{rd.2}\changeD{D2}As prior work points out~\cite{brooks1999dynamically}, 
static compiler analyses \emph{cannot} identify the bit-precision of dynamically allocated and initialized data arrays. \revD{We investigated several prior compiler works~\cite{canesche2022polynomial,stephenson2000bidwidth,rodrigues2013fast,campos2012speed,cong2005bitwidth} that perform bit-width identification. However,  such works are limited to identifying the bit-precision of statically\omcrii{-}allocated variables.}} 
 
\paratitle{\gf{\omcrii{\emph{Limitation 2:}}~Throughput-Oriented Execution}} \gf{\omcrii{Existing} \gls{PuD} architectures favor throughput-oriented execution as DRAM parallelism can \sgi{partially} hide the activation latency in a \uprog. To further improve throughput, prior works~\omcrii{\cite{hajinazarsimdram, peng2023chopper,mimdramextended}} use DRAM's \gls{BLP} to 
\li~distribute \sgi{independent \uprogs} across DRAM banks~\cite{hajinazarsimdram}, or
\lii~parallelize data writing and \gls{PuD} computation of \emph{different} \sgi{\uprogs} targeting \emph{different} banks~\cite{peng2023chopper}. However, such approaches cannot reduce the latency of a \emph{single} \sgi{\uprog}.}

\begin{figure}[ht]
    \centering
    \includegraphics[width=\linewidth]{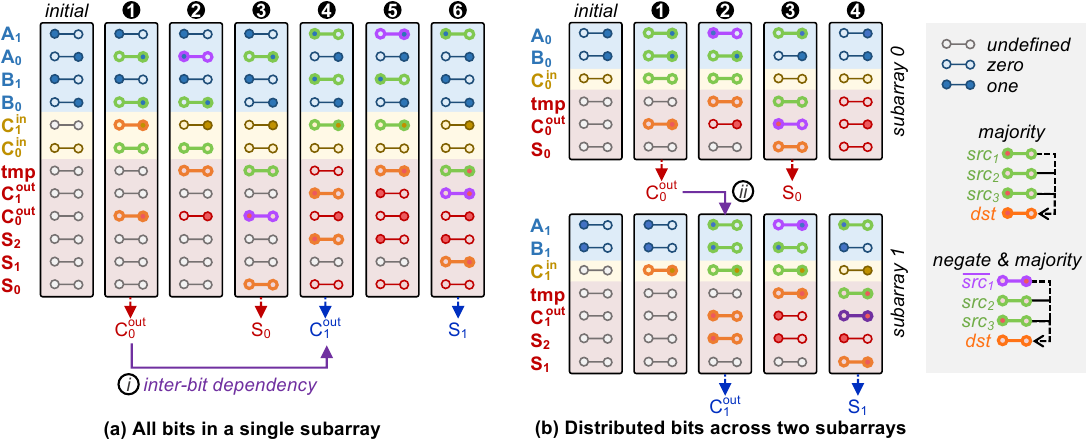}
    \caption{\gfcrii{Simplified bit-serial \gls{PuD} addition of two input arrays $A$ and $B$, each of which with two-bit data elements using (a)~one and (b)~two DRAM subarrays.}}
    \label{fig:uprogexample}
\end{figure}

\noindent \gf{\omcrii{\textbf{\emph{Opportunity 2:} DRAM Parallelism for Latency-Oriented Execution.}} 
\sgi{\revdel{We aim to leverage the internal parallelism of a DRAM bank to reduce the latency of a \emph{single} \uprog.}
We make the \emph{key observation} that several \gfasplos{primitives in} a \uprog \gfcrii{(\omcriii{i.e.,} \aap primitives that execute in-DRAM row copy or in-DRAM \texttt{MAJ3}/\texttt{NOT} operations)} can be executed concurrently, as they are independent of one another.
Fig.~\ref{fig:uprogexample} shows this opportunity for a two-bit addition.
\gfcrii{In conventional bit-serial execution (Fig.~\ref{fig:uprogexample}a), \emph{all} bits of the input arrays $A$ and $B$ are placed in a \emph{single} DRAM subarray.
Because of that, \emph{all} in-DRAM primitives in a \uprog are \emph{serialized}, enabling the execution of \omcriii{only} a single bit-position at a time.
In our example,  
\li~DRAM cycles \circled{1}--\circled{3} execute \texttt{MAJ3}/\texttt{NOT} operations over the \emph{least-significant bits} (LSBs) of the input arrays $A$ and $B$, i.e., $A_0$ and $B_0$; and afterwards
\lii~DRAM cycles \circled{4}--\circled{6} execute \texttt{MAJ3}/\texttt{NOT} operations over the \emph{most-significant bits} (MSBs) of the input arrays $A$ and $B$, i.e., $A_1$ and $B_1$.\footnote{\omcriii{For simplicity,} \gfcrii{we do \emph{not} depict DRAM cycles that perform in-DRAM row copy operations in Fig.~\ref{fig:uprogexample}.}}  
However, the only \emph{inter-bit} dependency in the \uprog is the \emph{carry propagation} (\circlediii{i} in Fig.~\ref{fig:uprogexample}a). 
In contrast, we can leverage \emph{bit-level parallelism} to \emph{concurrently} execute bit-independent in-DRAM primitives \omcriii{\emph{across multiple}} DRAM subarrays. 
In our example, we can reduce the overall latency of the bit-serial \gls{PuD} addition operation by 
\li~\emph{distributing} the individual bits of data-elements from arrays $A$ and $B$ across two DRAM subarrays (i.e., $subarray_0 \leftarrow  \{A_0,B_0\}; subarray_{1} \leftarrow  \{A_{1},B_{1}\}$), 
\lii~executing the required in-DRAM row copies (not shown) and \texttt{MAJ3}/\texttt{NOT} operations for the LSBs (DRAM cycles \circled{1}--\circled{3} in Fig.~\ref{fig:uprogexample}b) and MSBs (DRAM cycles \circled{2}--\circled{4} in Fig.~\ref{fig:uprogexample}b) \emph{concurrently}, and
\liii~serializing \emph{only} the carry generated from the LSBs to the MSBs of the input arrays (\circlediii{ii} in Fig.~\ref{fig:uprogexample}b).}}

\revdel{and how we could concurrently execute primitives for the addition if its operands were distributed across two different subarrays\revdel{ (compared to the serialization of all primitives in a single subarray, as existing \gls{PuD} architectures do currently)}\gfasplos{
, since} the only \omcrii{dependence} between the two bits of the addition is the \emph{carry propagation} (\circled{1} in Fig.~\ref{fig:uprogexample}a).
\gfasplos{By placing} each bit of the operand into a different subarray, we can parallelize independent primitives and serialize \emph{only} for the carry propagation (\circled{2} in Fig.~\ref{fig:uprogexample}b).}
\sgdel{We aim to leverage DRAM's internal parallelism to reduce the latency of a single \gls{PuD} operation. We make the \emph{key observation} that several row copies and TRAs in a \uprog can be executed in parallel. Fig.~\ref{fig:uprogexample} illustrates such observation.
The figure depicts the graph representation of the \uprog for a two-bit bit-serial addition operation and the execution time diagram of such \uprog using one (Fig.~\ref{fig:uprogexample}a) and two (Fig.~\ref{fig:uprogexample}b) DRAM subarrays. 
In the conventional sequential execution, all row copies and TRAs in the \uprog are serialized inside a single DRAM subarray, executing the required operations for the first-bit of the input operands first and then the second-bit of the input operand. However, the only data dependency between the operations for the first-bit and second-bit in the \uprog is the \emph{carry propagation} (\circled{1} in Fig.~\ref{fig:uprogexample}a). Therefore, by using two DRAM subarrays for computation, we can parallelize independent operations and \emph{only} serialize the carry propagation (\circled{2} in Fig.~\ref{fig:uprogexample}b), reducing the overall latency of the \gls{PuD} operation.} }

\gfcrii{Besides reducing the latency of bit-serial \gls{PuD} operations, carefully distributing individual bit positions across different DRAM subarrays within a DRAM bank enables the efficient realization of latency-friendly \emph{bit-parallel} \gls{PuD} arithmetic operations.
By mapping each bit position of a data element to a distinct subarray, our \gls{PuD} substrate can \emph{concurrently} perform \omcriii{bitwise} operations across all bits of the operand, thereby fully exploiting the parallelism inherent to bit-parallel arithmetic algorithms.}

\paratitle{\gf{\omcrii{\emph{Limitation 3:}}~High-Precision Computation}} 
\sgi{\gls{PuD} suffers from high latency for high bit-precision operations.}
\gf{\sgdel{\gls{PuD} suffers from \gfii{high latency that undermines the potential benefits of \gls{PuD} when implementing} key operations using high bit-precision.}%
For example, even when employing multiple (i.e., 16) \sgi{parallel} DRAM banks, SIMDRAM's throughput for \sg{32-bit and} 64-bit  division is 0.8$\times$ and 0.5$\times$ that of a \sg{16-core} CPU system~\cite{hajinazarsimdram}. \sgi{This is because} the latency of bit-serial multiplication and division scales \emph{quadratically} with the bit-precision. \sgdel{Such high latency directly impacts the performance of the \gls{PuD} substrate when executing multiplication/division-heavy workloads, leading to performance loss in such cases~\cite{hajinazarsimdram}.}}

\noindent \gf{\omcrii{\textbf{\emph{Opportunity 3:} Alternative Data Representation for High-Precision Computation.}} The high latency associated with high-precision computation is an \emph{inherent} property of coupling the binary numeral system with bit-serial computation. \omcrii{We} investigate \gfasplos{an} alternative data representation\gfasplos{, i.e.  the \emph{redundant binary representation (RBR})~\gfcrii{\cite{guest1980truth,phatak1994hybrid,lapointe1993systematic, olivares2006sad, olivares2004minimum}},} for high-precision computation.
\sgi{\gls{RBR} is a positional number system where each \gfcrii{bit-position}~$i$, which encodes $2^i$, is represented by two bits that can take on a value $v \in \{-1, 0, 1\}$, such that the magnitude of \gfcrii{bit-position}~$i$ is $v \times 2^i$.
For example, the 4-\gfcrii{bit-position} number \texttt{<0,1,0,-1>} represents $2^2 - 2^0 = 3$.
\gls{PuD} execution can take advantage of two \omcrii{key} properties of \gls{RBR}-based arithmetic:
\li~the operations no longer need to propagate carry bits through the full width of the data
(e.g., \gls{RBR}-based addition limits carry propagation to \emph{at most} two places~\sg{\cite{brown2002using}}), and
\lii~the operation latency is \emph{independent} of the bit-precision.}
\sgdel{, a positional numeral system that uses more bits than needed (e.g., two bits) to represent a single binary number.
In a \gls{RBR}-based system, each digit can take on any value from the set \texttt{\{-1, 0, 1\}}. Because there are three possible digit values, two bits are required to encode each digit. In conventional unsigned binary format, the $i^{th}$ digit represents $2^i$ multiplied by 0 or 1. In \gls{RBR}, the $i^{th}$ digit represents $2^i$ multiplied by \texttt{-1}, \texttt{0}, or \texttt{1}. An $n$-digit \gls{RBR} number $X = x_{n-1},x_{n-2}, ..., x_0$, where $x_i \in $  \texttt{\{-1, 0, 1\}} represents the value $\sum_{i=0}^{n-1} x_{i}2^{i}$. For example, the 4-digit number \texttt{<0,1,0,-1>} represents $2^2 - 2^0 = 3$. A \gls{RBR} number $X$ is often represented by two sets of bits, $X^{+}$ and $X^{-}$, representing the positive and negative powers of 2, respectively, that are added to the computed value. In the previous example, $X^{+} = $ \texttt{<0,1,0,0>} and $X^{-} = $ \texttt{<0,0,0,1>}. The two's complement of $X$ can be computed by subtracting $X^{-}$ from $X^{+}$ in the two's complement number system.
There are two main properties of RBR arithmetic that can benefit \gls{PuD} execution. 
First, using RBR for arithmetic operations (e.g., addition~\omcrii{\cite{wang2004new,jose2006delay}}) is almost \emph{carry-free}, i.e., there is no need to propagate carry bits through the full width of the addition operation.\footnote{\gf{\gls{RBR}-based addition limits carry propagation to \emph{at most} two digits~\sg{\cite{brown2002using}}.}} Second, the latency of an arithmetic operation is \emph{independent}. 
}}

\paratitle{Goal} \gf{Our \emph{goal} in this work is to mitigate the three limitations of \gls{PuD} architectures \omcriii{that arise due to the naive use of} a bit-serial execution model. 
To do so, we aim to \emph{fully} exploit the opportunities that DRAM's internal parallelism and dynamic bit-precision can provide to \gfcrii{reduce} the latency \gfcrii{and energy} of \gls{PuD} operations\gfcriii{. 
Concretely, we aim to} \gfcrii{\gfcriii{enable} 
\li~adaptive data-representation formats (two's complement and \gls{RBR}) for \gls{PuD} operands and 
\lii~flexible execution of different arithmetic algorithm implementations} (bit-serial and bit-parallel) \gfcrii{for \gls{PuD} instructions}.}  

\section{\prop Overview}
\label{sec:overview}

\begin{figure*}[!t]
    \centering
    \includegraphics[width=0.9\linewidth]{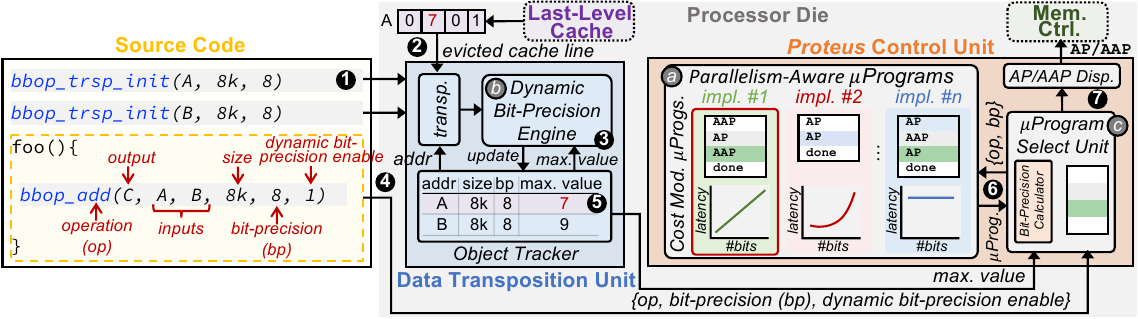}
    \caption{\gfcrii{\hl{Overview of the \mbox{\prop} framework.}}} 
     \label{fig:high-level}
\end{figure*}

{\gf{\prop is a \omcrii{new data-aware runtime} \gls{PuD} framework \gfcrii{that \emph{dynamically} adjust the bit-precision and, based on that, \emph{chooses} and \emph{uses} the most appropriate data representation and arithmetic algorithm implementation for a given \gls{PuD} operation.}
The \emph{key ideas} behind \prop are: 
\li~to \gfcrii{\emph{reduce} the bit-precision for \gls{PuD} operations by dynamically leveraging \emph{narrow values};
\lii~to \emph{parallelize} the execution of \emph{independent} in-DRAM primitives (i.e., \aaps in a \uprog) by leveraging \gls{SLP}~\cite{kim2012case} combined with \emph{bit-level parallelism}; and
\liii~to \emph{\omcriii{use}} alternative data representations (i.e., \gls{RBR}~\cite{guest1980truth,phatak1994hybrid,lapointe1993systematic, olivares2006sad, olivares2004minimum}) for high-precision computation.}

\gf{Fig.~\ref{fig:high-level} provides a high-level overview of \prop' framework, its main components, and execution flow. \prop is composed of three main components: 
\omcrii{
\li~\uproglib, 
\lii~\dynengine, and 
\liii~\uprogunit}.  
These components are implemented in hardware \omcrii{inside the} DRAM memory controller. The \uproglib and \uprogunit are part of \emph{\prop \gfcrii{C}ontrol \gfcrii{U}nit}. }
}
\gfcrii{In this section, we provide a high-level overview of \prop' main components (\cref{sec:overview:components}) and its execution flow (\cref{sec:overview:execution}). 
Detailed information on the implementation of each component is provided in~\cref{sec:implementation}. }



\subsection{Main Components of \prop}
\label{sec:overview:components}

\paratitle{\gf{Parallelism-Aware \uprog Library} \gfcrii{(\cref{sec:implementation:library})}}
\sgi{\prop incorporates a \uproglib  (\gfcrii{\circledii{a}} in Fig.~\ref{fig:high-level}) that consists of
\li~hand-optimized implementations of different \uprogs for key \gls{PuD} operations (each with different performance and bit-precision trade-offs), and
\lii~\costmodel.
For each operation, we implement multiple \uprogs (\cref{sec:implementation:library:optimization}) that use different
\li~bit-serial or bit-parallel algorithms and
\lii~data representation formats (i.e., two's complement or \gls{RBR}).
Each \uprog uses a novel data mapping that enables the \emph{concurrent} execution of multiple independent primitives across bits (\cref{sec:implementation:library:obps}).
The performance of each \uprog depends on the bit-precision, and
the \costmodel selects the best-performing \uprog for a given operation and target bit-precision.
}
\gf{\sgdel{\prop' \uproglib (\gfcrii{\circledd{a}} in Fig.~\ref{fig:high-level})  provides a range of optimized \uprogs for a given \gls{PuD} operation, each of which with different trade-offs in terms of performance and bit-precision. 
The \uproglib consists of 
\li~hand-tuned implementations of different \uprogs targeting key arithmetic operations, and
\lii~\costmodel.
First, to build a robust set of \gls{PuD} operations, we manually implement \uprogs using different
\li~bit-serial (e.g., ripple-carry adder) and bit-parallel 
\sgdel{(e.g., carry-select adder~\cite{bedrij1962carry}, Kogge–Stone adder~\cite{kogge1973parallel}, Ladner-Fischer adder~\cite{ladner1980parallel})}%
algorithms and 
\lii~data representation formats (i.e., two's complement and \gls{RBR}).  
We optimize each \uprog by 
\li~distributing input operands across different DRAM subarrays using an \gls{OBPS} data mapping, where bits from an input operand are scattered across DRAM subarrays; and 
\lii~parallelizing the execution of data-independent operations in a \uprog across different DRAM subarrays using \gls{SLP}~\cite{kim2012case} (to concurrent\juan{ly{}} execute \aaps in several subarrays) and LISA~\cite{chang2016low} (to transfer partial results across subarrays).
In the end, the \uproglib contains a collection of possible implementations of a desired \gls{PuD} operation, each with different time complexities that depend on the \gls{PuD} bit-precision. 
Second, the \costmodel aims to select the best-performing \uprog for a given operation and target bit-precision.}%
The \costmodel comprises of \prelut, which \sgi{list the most-suitable \uprog for each} bit-precision, and \emph{Select Logic} to identify the target \gls{LUT} for a \emph{bbop} instruction. 
\sgi{We \gfcrii{\emph{empirically} measure the throughput and energy efficiency of \uprogs in \uproglib while scaling the target bit-precision to} populate the \prelut.}
}

\paratitle{\gf{Dynamic Bit-Precision Engine} \gfcrii{(\cref{sec:implementation:dynamic-precision})}} \gf{\prop' \dynengine (\gfcrii{\circledii{b}} in Fig.~\ref{fig:high-level}) aims to identify the dynamic range of \emph{memory objects} associated with a \gls{PuD} operation.  
To do so, we dynamically identify \gfcrii{(in hardware)} the \emph{largest} input operand a \gls{PuD}'s memory object stores. In state-of-the-art \gls{PuD} architectures~\cite{hajinazarsimdram, peng2023chopper,gao2019computedram,ali2019memory,angizi2019graphide}, cache lines belonging to a \gls{PuD}'s memory object need to be transposed from the traditional horizontal data layout to a vertical data layout \emph{prior} to the execution of a \gls{PuD} operation. To efficiently perform such data transformation, SIMDRAM~\cite{hajinazarsimdram} implements a \emph{Data Transposition Unit}, which hides the data transposition latency by overlapping cache line evictions and data layout transformation. 
The \omcrii{\emph{Data Transposition Unit}} consists of an \emph{Object Tracker} table (a small cache that keeps track of memory objects that are used by \gls{PuD} operations) and \emph{Data \omcrii{Transposition} Engines}. The user/compiler informs the \emph{Object Tracker} of \gls{PuD}'s memory objects (both inputs and outputs) using a specialized instruction called \texttt{bbop\_trsp\_init}.
\prop leverages such \omcrii{a} \emph{Data Transposition Unit} to \omcrii{dynamically} identify \omcrii{in hardware} the largest value in a \gls{PuD}'s memory object by adding:
\li~a \gfcrii{\emph{new field}} in the \emph{Data Transposition Unit} called \emph{maximum value}, which stores the largest value in a given memory object\revdel{ \sg{evicted from a cache line}}; and
\lii~a \dynengine, which \gfmicro{scans} \omcrii{the} data elements of evicted cache lines, identifies the largest data value across all data elements and updates the stored \emph{maximum value} entry in the \emph{Data Transposition Unit}.\revdel{when necessary.}} 

\paratitle{\gf{\uprog Select Unit} \gfcrii{(\cref{sec:implementation:control-unit})}}  \gf{\prop' \uprogunit (\gfcrii{\circledii{c}} in Fig.~\ref{fig:high-level}) identifies the appropriate bit-precision for a \gls{PuD} operation based on \omcrii{the operation's} input data.
The \uprogunit has of a \li~\bitprec, which evaluates the target bit-precision based on the input operands of the \gls{PuD} operation and their associated maximum values, and
\lii~buffers to store the selected \uprog. 
}

\subsection{Execution Flow}
\label{sec:overview:execution}

\gf{\prop works in five} \gf{main steps.
In the first step (\omcrii{\circled{1}} in Fig.~\ref{fig:high-level}), the programmer/compiler utilizes specialized instructions (i.e., \omcriii{\texttt{bbop\_trsp\_init}}) to 
\li~register \omcrii{in} the \emph{Object Tracker} the address, size, and initial bit-precision \gfcrii{for each memory object used as an input, output, or temporary operand in a \gls{PuD} operation}; and
\lii~execute \gfcrii{a} \gls{PuD} \gfcriii{operation} over previously-registered memory objects. 
When issuing \gfcriii{an arithmetic} \bbop instruction, the programmer/compiler indicates whether or not dynamic bit-precision is enabled or disabled \omcrii{for that \bbop instruction}. \revD{\label{rd.2}\changeD{D2}When dynamic bit-precision is disabled, \prop' \dynengine is turned off, and the \uprogunit utilizes the user-provided bit-precision for \gfcrii{the} upcoming \gls{PuD} \gfcrii{operation related to the issued \bbop instruction}.\revdel{ This allows \prop to avoid performing redundant computations if the user has domain expertise and knows the dynamic range of \gls{PuD} operations beforehand.}}
In the second step, \sgi{if} the \dynengine is enabled, it intercepts evicted cache lines belonging to previously registered memory objects (\gfcrii{\circled{2}}) and identifies the largest value stored in the cache line. \sgi{If} the identified value is \emph{larger than} the current maximum \omcrii{value} stored in the \emph{Object Tracker}, the \dynengine updates the \emph{Object Tracker} with the up-to-date value (\omcrii{\circled{3}}). 
\gfcrii{The second step is repeated for all cache lines belonging \omcriii{to the} memory objects \omcriii{registered} in the \emph{Object Tracker}}. 
As in SIMDRAM~\cite{hajinazarsimdram}, our system employs lazy allocation \gfcrii{and maintains data coherence for \gls{PuD} memory objects through cache line flushing, using the \texttt{clflush} instruction~\cite{guide2016intel}.}
Thus, all memory objects initially reside within the CPU caches, and prior to PUD execution, all cache lines belonging to a \gls{PuD} operation \omcrii{are} evicted to DRAM, \gfmicro{which allows} \prop' \dynengine \gfmicro{to} access \emph{all} data elements of a \gls{PuD} operation prior to computation.\footnote{\gfasplos{\omcrii{Some real}-world \gls{PnM} architectures~\omcrii{\cite{upmem,devaux2019true,gomez2022benchmarking,gomez2021benchmarkingcut, gomezluna2021benchmarking,gomez2023evaluating}}} employ a similar execution model, where \emph{all} inputs need to be copied to \gls{PnM} cores \emph{prior} to \gls{PnM} execution.}} 
In the third step, the host CPU dispatches \gfcriii{the} \gfcrii{arithmetic} \bbop \gfcriii{instruction} (\omcrii{\circled{4}}) to \prop' Control Unit.
In the fourth step, \prop' \emph{Control Unit} receives the \bbop instruction from the CPU and the maximum values from the \dynengine (\omcrii{\circled{5}}), which are used as \omcrii{inputs} to the \uprogunit. Based on \omcrii{this} information, the \bitprec computes the target bit-precision and probes the \uproglib (\omcrii{\circled{6}}), which returns the best-performing \uprog and data format representation for the target \gfcrii{\gls{PuD}} operation \omcrii{and} \gfcrii{its associated} bit-precision.
In the fifth step, the \uprogunit dispatches the sequence of \aaps in the selected \uprog to DRAM (\omcrii{\circled{7}}).
When the host CPU reads back \gls{PuD} memory objects (not shown in the \juan{f}igure), 
\prop 
\li~performs the necessary data format conversions either 
from the reduced bit-precision to the user's specified bit-precision or from \gls{RBR} to two's complement \gfasplos{(thus maintaining system compatibility), and
\lii~\gfasplos{prepares the \dynengine for future accesses by resetting the current maximum data value stored in the \emph{Object Tracker}}}.  
} 
\section{\prop Implementation}
\label{sec:implementation}

\revdel{In this section, we describe the implementation details of \prop' main components. We first describe \prop' underlying subarray organization (Section~\ref{sec:implementation:subarray}). Then, we discuss the design of \prop'
\li~\uproglib (Section~\ref{sec:implementation:library}),
 \lii~\dynengine (Section~\ref{sec:implementation:dynamic-precision}), and
 \liii~\uprogunit (Section~\ref{sec:implementation:control-unit}).
}
\subsection{Subarray Organization}
\label{sec:implementation:subarray}

\begin{figure}[!htp]
    \centering
    \includegraphics[width=0.45\linewidth]{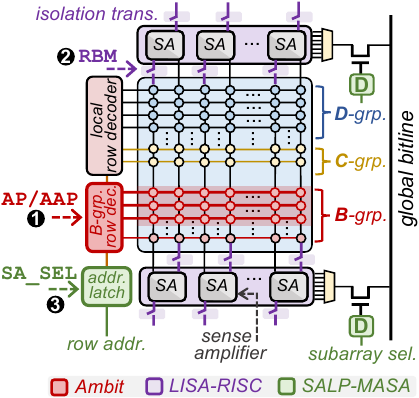}
    \caption{\gfcrii{\prop' subarray organization.}}
    \label{fig:implementation:subarray}
\end{figure}

\gfcrii{To efficiently perform \gls{PuD} operations, \prop uses a subarray organization that incorporates additional functionality to perform 
\li~in-DRAM logic primitives (i.e., \aaps), 
\lii~inter-subarray data copy, and 
\liii~subarray-level parallelism (\gls{SLP})~\cite{kim2012case}. 
Fig.~\ref{fig:implementation:subarray} illustrates the internal organization of a subarray in \prop, which is replicated across \emph{all} subarrays in a \prop-enabled DRAM bank.}

\paratitle{Performing Logic Primitives with Ambit}
\gf{%
\sgi{\prop reuses the subarray organization of Ambit~\cite{seshadri2017ambit} and SIMDRAM~\cite{hajinazarsimdram} \omcrii{(shown in Fig.~\ref{fig:implementation:subarray})} to enable logic primitive execution with only \omcrii{small} subarray modifications\revdel{ (a small row decoder that can activate up to three rows simultaneously)}. DRAM rows are divided into three groups:}
\sgdel{To enable the execution of logic primitives, \prop follows the same subarray organization of Ambit~\cite{seshadri2017ambit} and SIMDRAM~\cite{hajinazarsimdram}, requiring only minimal modifications to the DRAM subarray (namely, a small row decoder that can activate three rows simultaneously).
Like Ambit~\cite{seshadri2017ambit}, \prop divides DRAM rows into \emph{three groups}:}
\li~the \textbf{D}ata group (D-group), containing regular rows that store program data;
\lii~the \textbf{C}ontrol group (C-group), containing two rows pre-initialized with all-`0' and all-`1' values\revdel{\footnote{\gf{Ambit uses \texttt{AP} command sequences to implement Boolean AND and OR operations by simply setting one of the inputs (e.g., $C$) to 1 or 0. The AND operation is computed by setting $C$ to 0 (i.e., \texttt{MAJ(A, B, 0) = A AND B}). The OR operation is computed by setting $C$ to 1 (i.e., \texttt{MAJ(A, B, 1) = A OR B}).}}}; and 
\lii~the \textbf{B}itwise group (B-group), containing six rows (called \emph{compute rows}) to perform bitwise operations. 
The B-group rows are all connected to a special row decoder that can simultaneously activate \label{rd.5}\changeD{D5}\revD{three rows when performing an \texttt{AP} and two when performing an \texttt{AAP}} (\circled{1} in Fig.~\ref{fig:implementation:subarray}). }

\paratitle{Inter-Subarray Data Copy with LISA}
\sgi{\prop leverages LISA-RISC~\cite{chang2016low},
which dynamically connects adjacent subarrays using isolation transistors\revdel{ to copy an entire DRAM row}, to propagate intermediate data\revdel{ (e.g., carry)} \gfmicro{across subarrays}.
LISA-RISC \gfmicro{works} in four steps:
\li~activate the source row in the source subarray (latency: $t_{RAS}$);
\lii~use the LISA \emph{row buffer movement} command \omcrii{(\texttt{RBM}, \circled{2} in Fig.~\ref{fig:implementation:subarray})} to turn on isolation transistors, which copies data from the source subarray's local row buffer (LRB) to the destination subarray's LRB (latency: $t_{RBM}$, \SI{5}{\nano\second}\revdel{ in SPICE simulations}~\cite{chang2016low});
\liii~activate the destination row, to save the contents of the destination LRB into the \gfcrii{destination} row  (latency: $t_{RAS}$); and
\liv~precharge the bank (latency: $t_{RP}$).
\gfmicro{D}ue to DRAM's open bitline architecture~\cite{lim20121,takahashi2001multigigabit},
each LRB stores half of the row, so we must perform steps \lii--\liv\ twice to copy both halves of the row.}
\sgdel{\gf{To enable fast propagation of intermediate data (e.g., carry-propagation) across DRAM subarrays, \prop leverages LISA~\cite{chang2016low}. LISA creates a datapath between adjacent subarrays by connecting their
local row buffers using isolation transistors, which are controlled via a new DRAM operation called \emph{row buffer movement} (\texttt{RBM}, \circled{2} in Fig.~\ref{fig:implementation:subarray}). 
Such newly added datapath allows LISA to implement \underline{r}apid \underline{i}nter-\underline{s}ubarray bulk data {c}opy (i.e., LISA-RISC) using \sg{four} main steps \sg{(and associated DRAM timing parameters)}:
\li~activate \sg{($t_{RAS}$)} a source row in a subarray;
\lii~rapidly transfer the data in the activated source row buffers to the destination subarray's row buffer via the link created by LISA's isolation transistors \sg{($t_{RBM}$)};
\liii~activate \sg{($t_{RAS}$)} the destination row, which enables the contents of the destination row buffer to be latched into the destination row;
\liv~\sg{precharge ($t_{RP}$) the DRAM bank for subsequent memory requests. Since during an \texttt{ACT} the sense amplifier stores \emph{half} the data for a DRAM row (due to the open bitline DRAM architecture~\cite{lim20121,takahashi2001multigigabit}), to copy an entire DRAM row, the LISA-RISC operation needs to perform a second sequence of \texttt{ACT}-\texttt{RBM}-\texttt{PRE} commands. Thus,} \gfisca{the latency of a LISA-RISC operation is of $3t_{RAS}\ +\ 2t_{RP}\ +\ 2t_{RBM}$.\footnote{\gfisca{According to \cite{chang2016low} SPICE simulation analysis, the latency of a \texttt{RBM} command (i.e., $t_{RBM}$) is of 5~ns.}}}}}

\paratitle{Enabling Subarray-Level Parallelism with SALP} \gf{To enable the concurrent execution of \gfcrii{bit-}independent \gfcrii{primitives} in a \uprog, \prop leverages \gls{SLP}~\cite{kim2012case}.
SALP-MASA \sgi{(\emph{\omcrii{M}ultitude of \omcrii{A}ctivated \omcrii{S}ubarrays})} allows multiple subarrays in a bank to be activated concurrently by 
\li~pushing the global row-address latch to individual subarrays, 
\lii~adding a \sgi{designated}-bit latch (\textbf{D} in Fig.~\ref{fig:implementation:subarray}) to each subarray to ensure that only a single subarray's row buffer is connected to the global bitline, and
\liii~routing a new global wire (called \emph{subarray select}), controlled by a new DRAM command (\texttt{SA\_SEL}, \circled{3} in Fig.~\ref{fig:implementation:subarray}), allowing the memory controller to set/clear \sgi{each designated-bit latch}.}


\subsection{Parallelism-Aware \uprog Library}
\label{sec:implementation:library}

\subsubsection{\gfisca{One-Bit Per-Subarray (OBPS) Data Mapping}} 
\label{sec:implementation:library:obps}
\begin{figure}[ht]
\centering
  \centering
 \includegraphics[width=\linewidth]{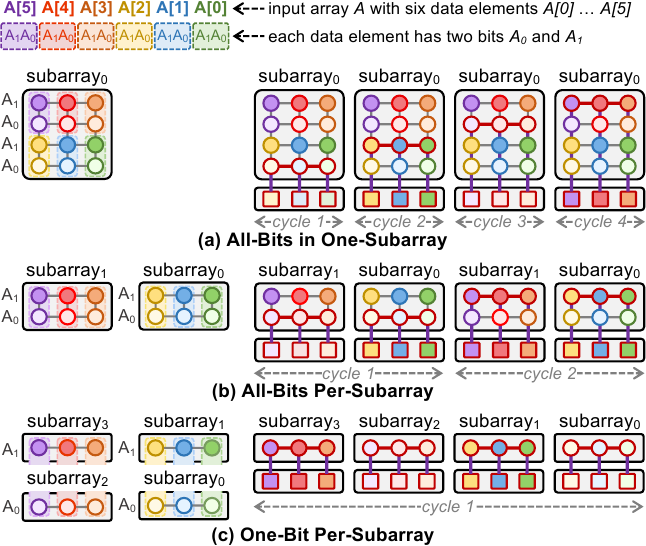}
  \caption{\gfcrii{Three data mappings for bit-serial computing.}}
  \label{fig:obps}
\end{figure}

To reduce the latency of \gls{PuD} operations (\cref{sec_motivation}), \prop employs a specialized data mapping called \emph{\gls{OBPS}}.
Bit-serial \gls{PuD} architectures can employ three data mappings, as Fig.~\ref{fig:obps} illustrates: 
\omcrii{\li~}\gls{ABOS}, 
\omcrii{\lii~}\gls{ABPS}, and 
\omcrii{\liii~}\gls{OBPS}. 
\gfcrii{Assume an example DRAM bank with four subarrays, a DRAM row size of three and an input array $A$ with six two-bit data elements.}

\gfcrii{First, the \emph{\gls{ABOS}} data mapping stores \emph{all} six two-bit data elements in \emph{one} DRAM subarray (Fig.~\ref{fig:obps}a). 
This data mapping limits the parallelism available for \gls{PuD} execution to that of a \emph{single} DRAM subarray, i.e., the number of DRAM columns simultaneously activated by a single \gls{PuD} primitive  (e.g., 65,536 DRAM columns per cycle in DDR4 memory \omcriv{chips}~\cite{standard2012jesd79}). 
In our example, the latency of executing a single \gls{PuD} primitive over \emph{all} data elements of the input array $A$ is four \gls{PuD} cycles \omcriii{(as shown in Fig.~\ref{fig:obps}a)}.\footnote{\gfcrii{We refer to a \emph{\gls{PuD} cycle} as the end-to-end latency required to execute a single \aap in-DRAM primitive}.}
Second, the \emph{\gls{ABPS}} data mapping distributes \emph{all} bits of multiple sets of the input array across \emph{multiple} DRAM subarrays (Fig.~\ref{fig:obps}b), allowing a \gls{PuD} primitive to execute concurrently on different portions of the input data stored in each subarray \omcriii{by} exploiting \emph{data-level parallelism}.
In our example, the latency of executing a single \gls{PuD} primitive over \emph{all} data elements of the input array $A$ while employing the \gls{ABPS} data mapping is two \gls{PuD} cycles \omcriii{(as shown in Fig.~\ref{fig:obps}b)}.
This is because, although execution across data elements can be parallelized by distributing them across multiple DRAM subarrays, the \gls{PuD} system must still serialize the execution of \gls{PuD} primitives across different bit positions of each data element, since \omcriii{\emph{all bits of a given data element}} are co-located within a single DRAM subarray under \gls{ABPS}.
Third, the \gls{OBPS} data mapping distributes \omcriii{each of} the $m$ individual bits of \omcriii{a given} data element of the input array to $m$ DRAM subarrays (Fig.~\ref{fig:obps}c), i.e., $subarray_0 \leftarrow  \{A_0\}, \dots$, $subarray_{m-1} \leftarrow  \{A_{m-1}\}$, allowing a \gls{PuD} primitive to execute concurrently on different bits of the input array stored in each subarray \omcriii{by} exploiting \emph{bit-level parallelism}.\footnote{If the number of subarrays is \omcriii{smaller} than the target bit-precision, \gls{OBPS} \emph{evenly} distributes the bits of input operands across the available subarrays.}  
In our example, the latency of executing a single \gls{PuD} primitive over \emph{all} data elements of the input array $A$ while employing the \gls{OBPS} data mapping is \omcriii{\emph{only}} a single \gls{PuD} cycle \omcriii{( as shown in Fig.~\ref{fig:obps}c)}.}


\subsubsection{\uprog Library Implementation}
\label{sec:implementation:library:optimization}

\gf{\prop leverages the subarray organization illustrated in Fig.~\ref{fig:implementation:subarray} and our \gls{OBPS} data mapping \omcrii{(Fig.~\ref{fig:obps})} to implement parallelism-aware \uprogs for key arithmetic operations \omcrii{(e.g., addition, multiplication)}.
We implement three classes of algorithms for arithmetic \gls{PuD} computations: \emph{bit-serial}, \emph{bit-parallel}, and \emph{RBR-based algorithms}. In \prop, each \uprog implementation 
\li~has an associated \gfcrii{\gbidx}, and 
\lii~is stored in a reserved memory space in DRAM (i.e., \emph{\uprog Memory}).} 

\paratitle{Bit-Serial Algorithms} \gf{We optimize \uprogs for bit-serial arithmetic operations (i.e., addition, subtraction, division, and multiplication)
\sgdel{in SIMDRAM}%
by concurrent\juan{ly} executing independent \aaps across different DRAM subarrays. Fig.~\ref{fig:uprogexample}b illustrates such a process for addition \gfcriv{using the \gls{OBPS} data mapping} (the process is analogous 
\juan{for} other arithmetic operations). \sgdel{In this example,}%
\prop implements a ripple-carry adder using majority gates in two main steps. 
\sgi{First},  \prop utilizes SALP-MASA to concurrent\juan{ly} execute the appropriate row copies and majority operations across $N$ different subarrays. 
\sgi{Second}, \prop utilizes LISA-RISC \sgdel{(and its associated \texttt{RBM} DRAM command)}%
to pipeline the carry propagation process (\gfcrii{\circledii{ii} in Fig.~\ref{fig:uprogexample}b)} from $subarray_i$ (e.g., $C_{out}^{0}$) to $subarray_{i+1}$ (e.g., $C_{in}^{1}$). This process repeats for all $N$ bits in the input operand. \prop reduces the latency of executing \omcriv{an $N$-bit} bit-serial addition from $8N + 1$ \aap cycles~\omcrii{\cite{hajinazarsimdram}} to $2N +7$ \aap cycles + $2(N-1)$ \texttt{RBM} cycles.\footnote{\label{ft.rc.7}\revC{To compute the number of \aap and \texttt{RBM} cycles in a \uprog, we implement each \uprog's algorithm using our cycle-\omcrii{level} data-accurate simulator (see~\cref{sec:methodology}\omcrii{; open sourced at~\cite{proteusgit}}). 
\revdel{Our simulator allows us to 
\li~compose a \uprog at the granularity of \aap and \texttt{RBM} operations and 
\lii~have individual control of the bits stored in a DRAM array while taking data modeling into consideration.}\revdel{ This means that, when simulating a TRA, our simulator actually computes the result of the majority operation for the data stored in the target DRAM rows.}  
We verified the correctness of a \uprog by testing it against several randomly generated data set combinations.}}  }

\paratitle{Bit-Parallel Algorithms}
\gf{We implement \sgi{bit-parallel variants of our \uprogs that}
leverage \emph{carry-lookahead logic} to decouple the calculation of the carry bits and arithmetic logic (e.g., addition). 
\sgi{Carry-lookahead logic can identify if any arithmetic on a bit will \emph{generate} a carry (e.g., both operands bits are `1' for an addition),
or if it will \emph{propagate} the carry value (e.g., only one operand bit for an addition is a `1').}
\sgdel{Carry-lookahead logic uses the concepts of \emph{generating} and \emph{propagating} carries. For example, in a binary addition, two input bits 
\li~\emph{generate} a carry only if \emph{both} bits are `1', and 
\lii~\emph{propagate} a carry only if there is an input carry.}%
\sgi{For $N$-bit operands, this} reduces time complexity compared to ripple-carry logic from $\mathcal{O}(N)$ to $\mathcal{O}(\log{}N)$, \gfcrii{where $N$ is the number of bits in the input operands.} We implement several carry-lookahead algorithms in \prop, including the carry-select~\cite{bedrij1962carry}, Kogge--Stone~\cite{kogge1973parallel}, Ladner--Fischer~\cite{ladner1980parallel}, and Brent--Kung~\cite{brent1982regular} adders, 
as building blocks to implement subtraction, multiplication, and division. 
Fig.~\ref{fig:paralleladder} \sgi{shows an example \prop implementation of a Kogge--Stone adder.}
\sgdel{illustrates \prop' implementation of an exemplary bit-parallel adder (i.e., the Kogge–Stone adder).}%


\begin{figure}[ht]
\begin{subfigure}{0.46\linewidth}
  \centering
 \includegraphics[width=\linewidth]{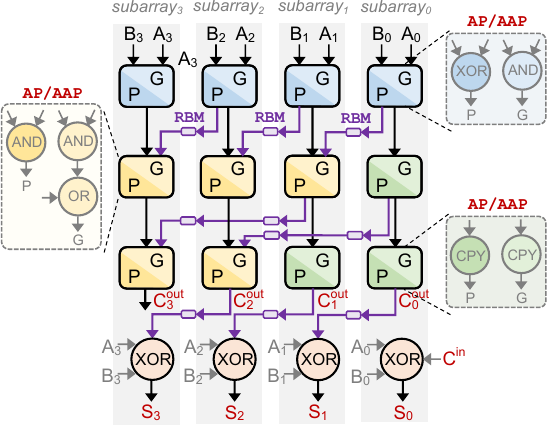} 
  \caption{4-bit Kogge–Stone adder}
  \label{fig:paralleladder}
\end{subfigure}
  \hfill 
\begin{subfigure}{0.46\linewidth}
  \centering
    \includegraphics[width=\linewidth]{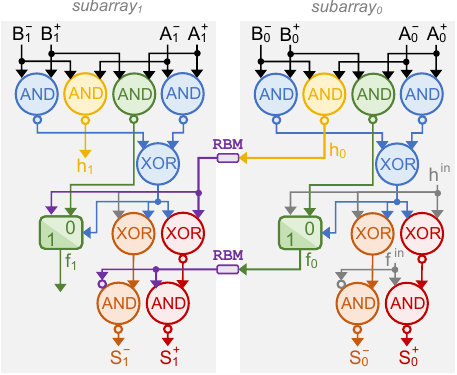}
 \caption{2-bit RBR adder}
  \label{fig:rbr_exemple}
\end{subfigure}
\caption{\prop' implementation of different adders. Bits $A_i$ and $B_i$ are stored \emph{vertically} in the same DRAM bitline \gfcrii{of subarray $i$ using the \gls{OBPS} data mapping}.}
\end{figure}

\revdel{The execution is divided into two main steps.}
In the first step, \prop performs \gfcrii{$2N +4$} inter-subarray data copies (using LISA-RISC) to copy the \emph{generate} and \emph{propagate} bits from $subarray_i$ to $subarray_{i+1}$.
In the second step, \prop performs a series of Boolean operations (using \aaps) to compute the next generate and propagate bits in parallel (using SALP-MASA) across \emph{all} DRAM subarrays.
These two steps repeat for $log(N)$ iterations. 
\gfcriv{The latency of executing \omcriv{an $N$-bit bit-parallel addition} using \prop is} $3log_2 N + 13$ \aap cycles + $2N + 4$ \texttt{RBM} cycles.
\gf{Even though the bit-parallel algorithms have a lower time complexity than the bit-serial algorithms, the \gfcrii{former} can require more inter-subarray copies\gfcrii{, i.e., $2N + 4$ \texttt{RBM} cycles for bit-parallel algorithms versus $2(N - 1)$  \texttt{RBM} cycles for bit-serial algorithms}.}  
}

\paratitle{RBR-Based Algorithms} \gfcrii{\gf{Fig.~\ref{fig:rbr_exemple} illustrates \prop' implementation of a two-bit \gls{RBR}-based adder~\cite{makino19968}. 
The adder operates in three steps. 
First, each digit $i$ generates an intermediate value $h_i$, computed \emph{only} from the corresponding input digit $i$. Second, the output value $f_i$ at digit $i$ is computed as a function of both the current digit and the preceding intermediate value $h_{i-1}$. 
Third, the $sum$ at digit $i$ depends on the current digit, $h_{i-1}$, and $f_{i-1}$. 
To propagate intermediate results between digits, \prop uses \texttt{RBM} commands to transfer the values of $h_i$ and $f_i$ from $subarray_i$ to $subarray_{i+1}$. 
The \gls{RBR}-based addition executes with a constant latency of 34 \aaps cycles and 8 \texttt{RBM} cycles. Beyond addition, \prop reuses the same \gls{RBR}-based adder design to support additional arithmetic operations, including subtraction and multiplication in the \gls{RBR} format.}}


\subsubsection{Cost Model Logic Implementation}
\label{sec:implementation:cost}

\begin{figure}[ht]
\centering
  \centering
 \includegraphics[width=0.65\linewidth]{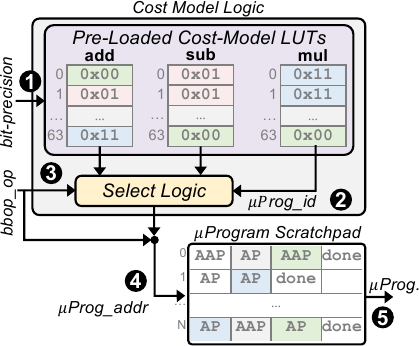}
  \caption{\omcrii{\prop' Cost Model Logic}.}
  \label{fig:cost-model}
\end{figure}

\gf{Fig.~\ref{fig:cost-model} depicts the hardware design of the \costmodel. 
The \costmodel has two main components: 
\li~one \gls{LUT} per \gls{PuD} operation, and
\lii~\emph{Select Logic}. 
Each \sgi{LUT row} represents a different bit-precision, and stores the \sgi{index} of the best-performing \uprog in the \sgi{library} for that \sgi{operation\omcrii{--}precision \omcrii{combination}}. 
\sgi{\gfcrii{We \emph{empirically} sized each} \gls{LUT} \gfcrii{to} contain 64 eight-bit rows (i.e., supporting up to 64-bit computation, and indexing up to 64 different \uprog implementations per \gls{PuD} operation)}. 
\revB{\label{rb.2}\changeB{B2}The \costmodel works in four main CPU cycles. It receives as input the \emph{bit-precision} (6 bits) and the \emph{bbop\_op} opcode (4 bits) of the target \gls{PuD} operation. 
In the first cycle, the \emph{bit-precision} indexes all the \glspl{LUT} in parallel (\circled{1} in Fig.~\ref{fig:cost-model}), selecting the best-performing \idx for the given bit-precision for all implemented \gls{PuD} operations (\circled{2}).
The \costmodel can \emph{quickly} query the \glspl{LUT} since they consist of a few (i.e., 16) small (i.e.,  64~B in size\revdel{; 64 rows, each holding eight-bit \idx}) SRAM arrays indexed in parallel. 
In the second cycle, based on the 4-bit \emph{bbop\_op} opcode, the \emph{Select Logic} chooses the appropriate \idx (\circled{3}).
In the third cycle, the \idx is concatenated with the \emph{bbop\_op} opcode to form the \gfcrii{\gbidx} (\circled{4}).
In the fourth \gfmicro{cycle}, the \gfcrii{\gbidx} indexes and fetches the best-performing \uprog from the \emph{\uprog Scratchpad} (\circled{5}). If the target \uprog is not loaded in the \emph{\uprog Scratchpad}, the \costmodel fetches it from the \emph{\uprog Memory} (not shown). }} 
\gfmicro{We estimate, using CACTI~\cite{cacti}, that the access latency and energy per access of the \SI{64}{\byte} SRAM array (used in our \costmodel)  is of  \SI{0.07}{\nano\second} (i.e., less than 1 CPU cycle) and \SI{0.00004}{\nano\joule}.\revdel{As a simple comparison point, the energy it takes to execute a 32-bit \gls{PuD} addition operation is 210.6~nJ~\cite{hajinazarsimdram}.}}


\subsubsection{\gfisca{Pareto Analysis}} 
\label{sec:implementation:pareto}
\gfisca{~We conduct a performance and energy Pareto \omcrii{analysis} to populate the \prelut. 
We model each \uprog using an analytical cost model that takes as input the target bit-precision, the number of elements used during computation, and the number of DRAM subarrays available. 
The analytical cost model outputs the throughput (in GOPs/s) and energy efficiency (in throughput/Watt) for each \uprog in the \uproglib.
We highlight our analyses for two main operations (i.e., addition and multiplication) since they represent linearly and quadratically-scaling \gls{PuD} operations, respectively. The analyses for subtraction and division follow similar observations.
In our analyses, we evaluate a SIMDRAM-like \gls{PuD} architecture using the three data mapping schemes described in Fig.~\ref{fig:obps}.
We assume a DRAM bank with 64 \omcrii{\gls{PuD}-capable} DRAM subarrays and a subarray with 65,536 columns.
We vary the number of input elements as multiples of the number of DRAM columns per subarray (from 1 DRAM subarray with \omcriv{64K} input elements to 64 DRAM subarrays with \omcriv{4M} input elements) for our measurements. 
} 

\begin{figure}[ht]
    \centering
    \includegraphics[width=\linewidth]{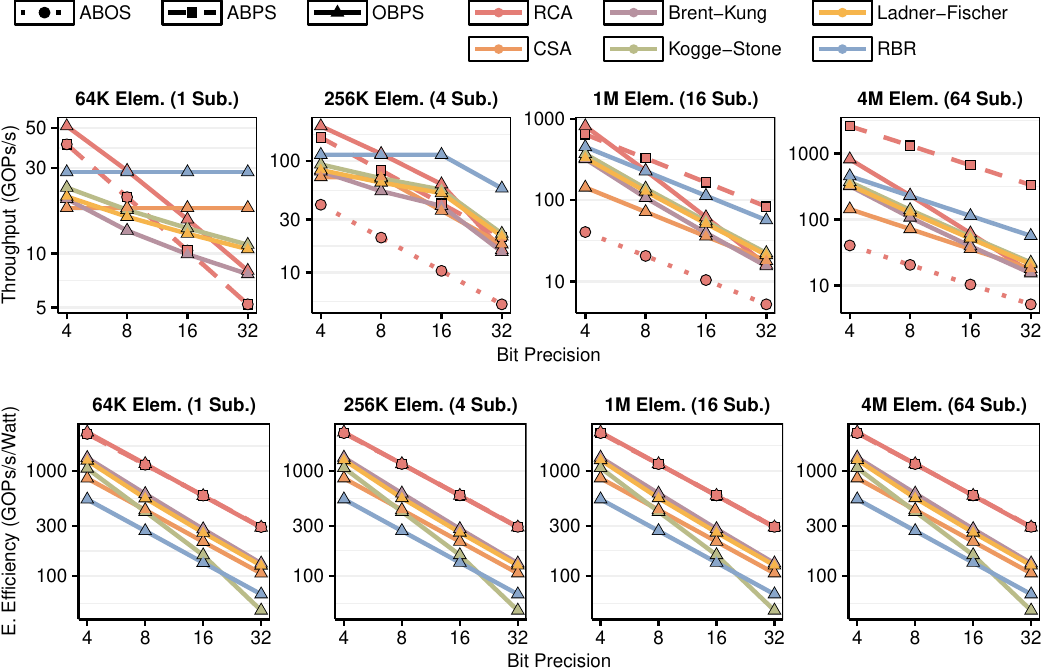}
    \caption{\omcrii{Pareto analysis for throughput (top) and energy efficiency (bottom) for \gls{PuD} addition operations. 
    Dotted lines represent \gls{ABOS};
    dashed lines represent \gls{ABPS}; straight lines represent \gls{OBPS} data mapping. }}
    \label{fig:pareto_addition}
\end{figure}

\paratitle{\gfisca{Linearly\omcrii{-}Scaling \gls{PuD} Operations}} \gfisca{Fig.~\ref{fig:pareto_addition} shows the throughput (\gfcriv{$y$-axis}; top) and energy efficiency (\gfcriv{$y$-axis}; bottom) of six \uprog implementations for a linearly\omcrii{-}scaling \gls{PuD} operation (i.e., \omcrii{integer} addition) \omcriv{for different bit-precision values ($x$-axis)}.
\gfcriv{Each subplot depicts the different input data sizes we use in our analysis.}
\gfcrii{For this analysis, we implement the following addition algorithms: 
ripple-carry adder (RCA), 
carry-select adder (CSA)~\cite{bedrij1962carry},
Brent-Kung adder~\cite{brent1982regular},
Kogge–Stone adder~\cite{kogge1973parallel}, Ladner-Fischer adder~\cite{ladner1980parallel}, using 
\gfcrii{\li~both} two's complement and \gls{RBR} \gfcrii{data format representations; and
\lii~\gls{ABOS}, \gls{ABPS}, and \gls{OBPS} data mappings}. 
Note that the bit-parallel adder can \emph{only} be implemented using the \gls{OBPS} data mappings.}
We make two observations. 
First, \omcrii{in terms of} throughput, the best-performing adder implementation varies depending on the target bit-precision and number of input elements.
\gfcrii{The achievable throughput ultimately depends on a combination of the number of \aaps that can be concurrently executed across DRAM subarrays and the number of inter-DRAM subarray operations required to implement the adder. 
In general, we empirically observe that as the input data size increases \gfcriv{(see subplots' titles)}, the number of inter-DRAM subarray operations also increases and eventually dominates the overall execution time.}
For \emph{small bit-precision} and \emph{small input size} (i.e., bit-precision smaller than 8, and \omcriv{fewer} than \omcriv{256K} input elements), the bit-serial RCA using the \gls{OBPS} data mapping provides 
\gfcrii{the highest throughput, while for \emph{large bit-precision} and \emph{small input size} (i.e., bit-precision larger than 8, and \omcriv{fewer} than \omcriv{256K} input elements), the \gls{RBR} adder using the \gls{OBPS} data mapping provides the highest throughput.}
For large-enough input sizes \gfcrii{(i.e., more than \omcriv{1M} input elements)}, employing the \gls{ABPS} data mapping leads to the highest throughput, independent of the bit-precision. 
This is because when more DRAM subarrays are involved in the execution of the target \gls{PuD} operation, the inter-subarray data transfers dominate overall execution time in the \gls{OBPS} implementations. 
Second, \omcrii{in terms of} energy efficiency, the bit-serial implementation of RCA provides the best throughput/Watt for \gls{ABOS}, \gls{ABPS}, and \gls{OBPS}, independent of the bit-precision and input size. 
This is because 
\li~the number of \aaps performed to execute RCA is the same \emph{independent} of the data mapping, and
\lii~the energy the bit-parallel algorithms consume is dominated by inter-subarray operations\gfcrii{, which is \emph{not} present in bit-serial implementations}.}

\begin{figure}[ht]
    \centering
    \includegraphics[width=\linewidth]{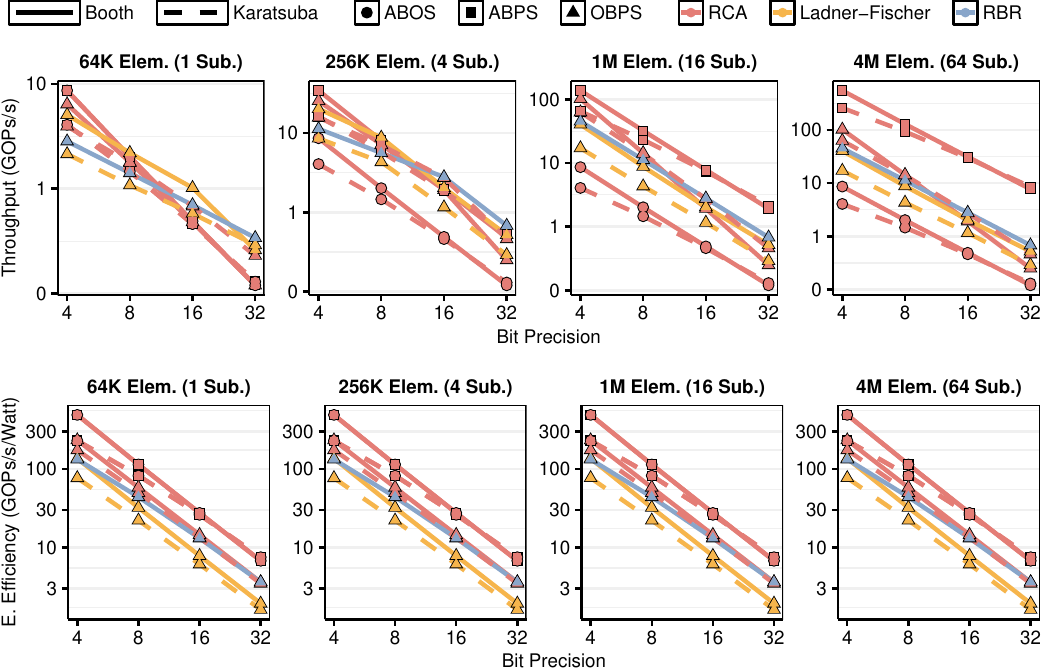}
    \caption{\omcrii{Pareto analysis for throughput (top) and energy efficiency (bottom) for multiplication. Straight lines represent the \gfcrii{Booth's} multiplication method~\cite{booth1951signed}; dashed lines represent the Karatsuba~\cite{karatsuba1962multiplication} multiplication method.  }}
    \label{fig:pareto_multiplication}
\end{figure}

\paratitle{\gfisca{Quadratically\gfcrii{-}Scaling \gls{PuD} Operations}} \gfisca{Fig.~\ref{fig:pareto_multiplication} shows the throughput (top) and energy efficiency (bottom) of six \uprog implementations for a quadratically\omcrii{-}scaling \gls{PuD} operation (i.e., \omcrii{integer} multiplication).
We implement \gls{PuD} multiplication \gfcrii{operations as a triplet composed of \li~the multiplication method (i.e., \gfcrii{Booth's multiplication algorithm~\cite{booth1951signed}} or the divide-and-conquer Karatsuba~\cite{karatsuba1962multiplication} multiplication);
\lii~different methods for addition (i.e., bit-serial RCA, bit-parallel Ladner-Fischer~\cite{ladner1980parallel}, and \gls{RBR}-based adder); and 
\liii~data mappings (i.e., \gls{ABOS}, \gls{ABPS}, and \gls{OBPS})}.
\gfcrii{Note that \gls{PuD} multiplication operations that use bit-parallel and RBR-based adders can \emph{only} be implemented using the \gls{OBPS} data mapping.}
We make two observations.
First, \omcrii{in terms of} throughput, the best-performing multiplier implementation varies depending on the bit-precision and number of input elements. 
For small bit-precision and small input size (i.e., bit-precision smaller than 8, and \omcriv{fewer} than \omcriv{64K} input elements), \gfcrii{Booth's} bit-serial multiplication \omcrii{with \gls{ABOS} data mapping} provides the highest throughput\gfcrii{, while for medium bit-precision and small input size (i.e., bit-precision from 8 to 16 and \omcriv{fewer} than \omcriv{64K} input elements), \gfcrii{Booth's} bit-parallel multiplication \omcrii{with the \gls{OBPS} data mapping} provides the highest throughput.}
For high bit-precision and small-to-medium input size (i.e., bit-precision larger than 32 and \omcriv{fewer} than \omcriv{256K} input elements), \gfcrii{RBR-based multiplication using \gls{OBPS} data mapping} provides the highest throughput.
\omcrii{For} large-enough input sizes (i.e., larger than \omcriv{1M} input elements), employing \gfcrii{Booth's bit-serial RCA-based multiplication using} \gls{ABPS} data mapping leads to the highest throughput, independent of the bit-precision. 
Second, \omcrii{in terms of} energy efficiency, \gfcrii{Booth's bit-serial RCA-based} multiplication implementation provides the best throughput/Watt for \gls{ABOS}, \gls{ABPS}, and \gls{OBPS}, independent of the bit-precision and input size\gfcrii{, since
\li~the number of \aaps required to execute the addition step is the same regardless of the data mapping and
\lii~the energy of the bit-parallel-based algorithms is dominated by the large number of inter-subarray operations they require.
}
}

\subsubsection{\gfisca{Non-Arithmetic \gls{PuD} Operations.}} 

\gfisca{We also equip \prop' \uproglib with SIMDRAM's implementations of non-arithmetic \gls{PuD} operations~\omcrii{\cite{hajinazarsimdram}}, including
\li~$N$-bit logic operations (i.e., \texttt{AND}/\texttt{OR}/\texttt{XOR} of more than two input bits), \lii~relational operations (i.e., equality/inequality check, greater than, maximum, minimum), 
\liii~predication, and \liv~bitcount and ReLU~\cite{goodfellow2016deep}.}

\subsection{Dynamic Bit-Precision Engine}
\label{sec:implementation:dynamic-precision}

\gf{The \dynengine comprises a simple reconfigurable $n$-bit comparator and a \gls{FSM}. For each evicted cache line, the \gls{FSM} probes the \emph{Object Tracker} and identifies if the \omcrii{incoming  evicted} cache line belongs to a \gls{PuD}'s memory object. If it does, the \gls{FSM} executes \gfmicro{four} operations.
First, it reads the bit-precision value (specified by the \texttt{bbop\_trsp\_init} instruction) and the current maximum value stored in the \emph{Object Tracker} for the given memory object.
Second, it uses the bit-precision value to configure the $n$-bit comparator.
Third, it inputs to the $n$-bit comparator all $n$-bit values in the \omcrii{incoming} cache line (one \omcrii{at a} time) and the current maximum value. 
Fourth, after all the $n$-bit values are processed, if any value in the \omcrii{incoming} cache line is larger than the current maximum value, the \gls{FSM} sends an update signal to the \emph{Object Tracker} alongside the new maximum value.}
\gfmicro{The energy cost of identifying the largest element in a \SI{64}{\byte} cache line is \SI{0.0016}{\nano\joule}~\cite{han2016eie}. That represents an increase in 0.084\% in the energy of an \gls{LLC} eviction~\cite{damov,muralimanohar2007optimizing,tsai:micro:2018:ams}, which \emph{needs} to happen prior to \gls{PuD} execution regardless.}

\subsection{\uprog Select Unit}
\label{sec:implementation:control-unit}

\revdel{\gf{The \uprogunit aims to identify and select the appropriate bit-precision and \uprog for a \gls{PuD} operation. It consists of a 
\li~\bitprec, which evaluates the target bit-precision based on the input operands of the target \gls{PuD} operation and their associated maximum values, and
\lii~buffer space to store the selected \uprog.}}

\paratitle{Calculating Bit-Precision} \gf{The \uprogunit needs to address two scenarios when calculating the bit-precision for \gls{PuD} operations: \emph{vector-to-vector} \gls{PuD} operations, and \emph{vector-to-scalar} reduction \gls{PuD} operations.
In \omcrii{\emph{vector-to-vector}}, the target \gls{PuD} operation implements a parallel \emph{map} operation, in which inputs and outputs are data vectors. For such operations, the bit-precision can be computed \emph{a priori}, using the maximum values the \dynengine provides, \emph{even} in the presence of chains of \gls{PuD} operations. In such a case, the \gfcrii{\emph{Bit-Precision Calculation Engine}} updates the \emph{Object Tracker} with the maximum possible output value for \emph{each} \gls{PuD} in the chain. 
For example, assume a kernel that executes \texttt{D[i]=(A[i]+B[i])$\times$C[i]} as follows:   }   

\begin{center}
\vspace{-10pt}
\tempcommand{.8}
  \resizebox{\columnwidth}{!}{
\begin{tabular}{ll}
\texttt{bbop\_add(tmp, A, B, 8k, 8, 1)};      & // tmp $\leftarrow$ A + B      \\
\texttt{bbop\_mul(D, tmp, C, 8k, 8, 1)};   & // D $\leftarrow$ tmp $\times$ C
\end{tabular}
}
\vspace{-10pt}
\end{center}

\noindent \gf{Assume that the maximum value of \texttt{A}, \texttt{B}, and \texttt{C} are $3$, $6$, and $2$, respectively. In this case, the \uprogunit
\li~computes the bit-precision for the addition operation as $\ceil*{\log_2 (3+6)} = 4\ bits$;
\lii~updates the \emph{Object Tracker} entry of \texttt{tmp} with the maximum value of the addition operation (i.e., $9$);
\liii~computes the bit-precision for the multiplication operation as $\ceil*{\log_2 (9\times2)} = 5\ bits$ \sg{using an $n$-bit scalar ALU};
\liv~updates the \emph{Object Tracker} entry of \texttt{D} with the maximum value of the multiplication (i.e., $18$).}

\gf{In \emph{vector-to-scalar} reduction, the \gls{PuD} operation implements a parallel \emph{reduction} operation, where the inputs are vectors and the output is a scalar value. In this case, the bit-precision \emph{cannot} be computed with \emph{only} the maximum input operands without causing \emph{overprovision\omcrii{ing}}, since in a reduction, each element contributes to the bit-precision of the scalar output. Therefore, for \emph{vector-to-scalar} reduction \gls{PuD} operations, the \uprogunit needs to 
\li~fetch from DRAM the row containing the carry-out bits produced during partial steps\footnote{\gf{\prop implements \gls{PuD} reduction operations using \emph{reduction trees}~\cite{mimdramextended}. Thus, a partial step refers to a level of the reduction tree.}} of the \gls{PuD} reduction;
\lii~evaluate \gfasplos{if} a partial step generated an overflow (i.e., check if any carry-out bit is `\texttt{1}'); and
\liii~increment the bit-precision for the next partial step \omcrii{if overflow is detected}.\revdel{ We design the \uprogunit to fully overlap the latency of bit-precision computation with \uprog execution.} }

\paratitle{Hardware Design} \gf{The \uprogunit comprises of simple hardware units:
\li~an \emph{$n$-bit ALU} to compute the target bit-precision, 
\lii~a \emph{Fetch Unit} to generate load instructions for carry re-evaluation, and
\liii~a \emph{\uprog Buffer} to store the currently running \uprog.}\revdel{\footnote{\label{ft.re.2}\revE{We do \emph{not} leverage \gls{PuD} operations to compute the target bit-precision for two reasons. 
First, the bit-precision information is required by \prop \emph{Control Unit} (placed within the memory controller) \emph{prior} to issuing \gls{PuD} operations for the target computation. 
Second, \prop calculates the target bit-precision for a given \gls{PuD} operation by identifying the largest data elements within the memory objects involved in the \gls{PuD} operation. This involves computing the target operation using at most two data elements. Offloading such computation to \gls{PuD} would incur wasted resources from the massively parallel \gls{PuD} system, high latency, and high energy cost.}}}

\subsection{\gfisca{\omcriv{Other Considerations}}}
\label{sec:implementation:alltogether}

\paratitle{\gfisca{Data Format Conversion}}
\gfisca{\revdel{Concretely, \prop might require 
\li~distributing the input operands stored vertically in one DRAM subarray to multiple DRAM subarrays (to implement the \gls{OBPS} data mapping) or
\lii~convert input operands from their two's complement representation to the RBR. 
\prop realizes both data format conversions using \gls{PuD} primitive operations.}
To distribute the bits of input operands to different DRAM subarrays, 
\prop issues \texttt{RBM} commands from the DRAM row storing bit $i$ in the source DRAM subarray to the target DRAM row in the destination DRAM subarray $i$, i.e., $row[dst]_{subarray_i} \leftarrow RBM (row[src]_i); i \in [1, bits]$. 
To convert the data stored in two's complement to its equivalent \gls{RBR} (see~\cref{sec_motivation}), \prop performs in-DRAM bitwise operations, as Table~\ref{table:data_conversion} describes.}

\begin{table}[ht]
\caption{Conversion from two's complement to RBR.}
   \centering
   \footnotesize
   \tempcommand{0.8}
   \resizebox{\columnwidth}{!}{
\begin{tabular}{r||ccc}
\toprule
\textbf{Input X $\rightarrow$}                       & \multicolumn{1}{c|}{2}       & \multicolumn{1}{c|}{-1}      & -7      \\ 
\textbf{two's complement $\rightarrow$}                & \multicolumn{1}{c|}{0 0 1 0} & \multicolumn{1}{c|}{1 1 1 1} & 1 0 0 1 \\ \midrule \midrule
\textbf{Steps to convert two's to RBR $\downarrow$}    & \multicolumn{3}{c}{\textbf{Output $\downarrow$}}                                  \\ \hline \hline
Extract most-significant bit (MSB)      & \multicolumn{1}{c|}{0}       & \multicolumn{1}{c|}{1}       & 1       \\ 
\gfcrii{buffer1}: broadcast MSB to all subarrays & \multicolumn{1}{c|}{0 0 0 0} & \multicolumn{1}{c|}{1 1 1 1} & 1 1 1 1 \\ 
\gfcrii{buffer2}: \texttt{NOT}(\gfcrii{buffer1})                   & \multicolumn{1}{c|}{1 1 1 1} & \multicolumn{1}{c|}{0 0 0 0} & 0 0 0 0 \\ 
~X + 1                                  & \multicolumn{1}{c|}{1 1 1 0} & \multicolumn{1}{c|}{0 0 0 1} & 0 1 1 1 \\ \hline 
$X-$ = \gfcrii{buffer1} \& (~X + 1)                 & \multicolumn{1}{c|}{0 0 0 0} & \multicolumn{1}{c|}{0 0 0 1} & 0 1 1 1 \\ 
$X+$ = \gfcrii{buffer2} \& (X)                      & \multicolumn{1}{c|}{0 0 1 0} & \multicolumn{1}{c|}{0 0 0 0} & 0 0 0 0 \\ 
\bottomrule
\end{tabular}
}
\label{table:data_conversion}
\end{table}

\paratitle{\gfisca{System Integration}} \gfisca{
\prop leverages the \emph{same} system integration solutions as in prior \gls{PuD} systems~\cite{hajinazarsimdram,mimdramextended}, including: 
\li~ISA extensions included to the host CPU ISA that the programmer utilizes to launch \emph{bbop} instructions;
\lii~a hardware control unit, alongside the memory controller, to control the execution of \uprogs; and
\liii~a hardware transposition unit, placed between the \gls{LLC} and the memory controller, to transpose data from the native horizontal data layout to the \gls{PuD}-friendly vertical data layout.
\revdel{As in~\cite{hajinazarsimdram,mimdramextended}, we assume that the operating system (OS) can provide support for data allocation and data mapping of operands in the DRAM bank dedicated for \gls{PuD} computing and the \gls{PuD} substrate operates directly on physical addresses.} 
\revdel{We acknowledge that such system integration assumptions are potential limitations of SIMDRAM and \prop (and several other \gls{PuD} architectures~\cite{ferreira2021pluto, ferreira2022pluto, li2017drisa, deng2018dracc}). 
We leave a more robust system integration implementation to future work.}
}

\paratitle{Limitation of \gls{SLP}} \gfisca{\gls{SLP} is limited by the $t_{FAW}$ DRAM timing constant~\cite{jedec2012ddr3, jedec2012jedec, jedec2017jedec}, which corresponds to the time window during which at most four \texttt{ACT} commands can be issued per DRAM rank. 
This constraint protects against the deterioration of the DRAM reference voltage. 
DRAM manufacturers have been able to
relax $t_{FAW}$ substantially in commodity DRAM chips~\cite{micron2013tfaw}, 
as well as to perform a targeted reduction of this parameter specifically for \gls{PIM} architectures where it becomes
a performance bottleneck~\cite{he2020newton,lee2021hardware}. These advances suggest that
\omcrii{$t_{FAW}$ likely does} \emph{not} limit \prop' scalability in commodity DRAM chips.}

\paratitle{\omcrii{Proteus for Floating-Point Operations}} \prop' dynamic bit-precision computation can be employed in two different stages of floating-point arithmetic for \gls{PuD} operations, during exponent and mantissa computation, \omcrii{using} three steps. \Copy{R1.1D}{\asplosrev{\hl{First, the \mbox{\dynengine} identifies the sign, exponent, and mantissa bits in floating-point numbers stored in the evicted cache lines of \mbox{\gls{PuD}} memory objects.
Second, the \mbox{\dynengine} updates the \emph{Object Tracker} with the maximum exponent and the mantissa values for the \mbox{\gls{PuD}} memory object, in case the \omcrii{identified} maximum values are smaller than the values stored in the current evicted cache line. This process is analogous to the execution flow described in \mbox{\cref{sec:overview:execution}} for integer operands, with the addition of the fields for the maximum exponent and maximum mantissa values in the \emph{Object Tracker}.
Third, \mbox{\prop} performs the target in-memory floating-point computation by issuing~\mbox{\cite{dualitycache}}
\li~bit-serial subtraction (addition) \mbox{\gls{PuD}} operations to calculate the resulting exponents, followed by 
\lii~bit-serial addition (multiplication) \mbox{\gls{PuD}} operations to calculate the resulting mantissas for a vector addition (multiplication) instruction~\mbox{\cite{dualitycache}}. Note that the bit-serial \mbox{\gls{PuD}} operations involved in \li--\lii~are vector-to-vector \mbox{\gls{PuD}} operations. As such, \mbox{\prop} can leverage the maximum exponent and mantissa values}\omrev{\hl{ stored in the \emph{Object Tracker}}}\hl{ to set the required bit-precision for such bit-serial operations, following the same approach described in~\mbox{\cref{sec:implementation:control-unit}}.}}}

\revdel{\Copy{R1.1E}{\asplosrev{\hl{We identify another opportunity to exploit narrow values during mantissa computation by identifying \emph{repeating patterns} in the mantissa to}\omrev{\hl{ configure}}\hl{ the bit-precision for the target operation. For example, when representing \texttt{+1.3} in floating-point (sign = `\texttt{0}', exponent = `\texttt{01111111}', mantissa = `\texttt{01001100110011001100110}'), the mantissa bits `\texttt{0011}' repeats five times. 
Based on that, we can perform the bit-serial mantissa computation by
\li~issuing \mbox{\gls{PuD}} operations (e.g., adding or multiplying the mantissa) for the first iteration of the repeating pattern `\texttt{0011}' and 
\lii~copying (using RowClone~\mbox{\cite{seshadri2013rowclone}}) the remaining bits of the output mantissa depending on the identified repeating pattern (i.e., four times in our example). 
However, to implement such}\omrev{\hl{ a}}\hl{ scheme, \mbox{\prop} would need to identify variable-length patterns across the mantissa bits for all data elements involved in the computation,}\omrev{\hl{ which would incur}}\hl{ non-trivial hardware complexity. Thus, we leave the realization of}\omrev{\hl{ this optimization}}\hl{ approach as future work.}}} 
}

\section{Methodology}
\label{sec:methodology}

\label{r3.1}\Copy{R3.1}{We implement \prop using an in-house cycle-\asplosrev{\hl{level}} simulator \gfcrii{(which we open-source at~\cite{proteusgit})} and compare it to a real multicore CPU (Intel Comet Lake~\cite{intelskylake}),
a real high-end GPU (NVIDIA A100 \asplosrev{\hl{using CUDA and tensor cores}}~\cite{a100}), and a \gfasplos{simulated} state-of-the-art \gls{PuD} framework  (SIMDRAM~\cite{hajinazarsimdram}). 
In our evaluations, the CPU code uses AVX-512 instructions~\cite{firasta2008intel}.
\asplosrev{\hl{Our simulator \omcrii{is} rigorously validated against SIMDRAM~\mbox{\cite{hajinazarsimdram}} and MIMDRAM~\mbox{\cite{mimdramextended}}'s gem5~\mbox{\cite{gem5}} implementation~\mbox{\cite{mimdramgit}}. The simulator 
\li~is cycle-level accurate}\omrev{\hl{ with regard to}}\hl{ DRAM commands
 and 
\lii~accounts for the data movement cost of cache line eviction on a per-cycle basis.}}
Our simulation accounts for the additional latency imposed by SALP~\cite{kim2012case} on \texttt{ACT} commands, i.e., the extra circuitry required to support SALP incurs an extra latency of \SI{0.028}{\nano\second} to an \texttt{ACT}~\cite{hassan2022case}, which is less than  0.11\% extra latency \omcrii{of} an \texttt{AAP}. 
To verify the functional correctness of our applications, our simulation infrastructure \omcrii{performs} functional verification over application's data when performing \gls{PuD} operations.
We did \emph{not} observe any difference from the \asplosrev{\hl{golden}} outputs.}
\gfcriv{We open-source our simulation infrastructure at \url{https://github.com/CMU-SAFARI/Proteus}.}

Table~\ref{table_parameters} shows the system parameters we use in our evaluations.
\revdel{\changeA{A1}\revA{\label{ra.1}We evaluate SIMDRAM and \prop execution time by isolating the main kernel that SIMDRAM/\prop executes (i.e., the offloaded \gls{PuD} instructions) and evaluating its performance. 
We estimate the end-to-end speedup SIMDRAM/\prop provides for each application by applying Amdahl's law~\cite{amdahl1967validity}. 
Thus, SIMDRAM and \prop' end-to-end speedup is given by: $((1 - kernel\_time) + \frac{kernel\_time}{PUD\_impro})^{-1}$; where $PUD\_impro$ is the speedup SIMDRAM/\prop provides for the offloaded kernel compared to the CPU execution, and $kernel\_time$ is the percentage of the total execution time the offloaded kernel represents when executing the application on the baseline multi-core CPU.}}
\gfisca{To measure CPU energy consumption, we use Intel RAPL~\cite{hahnel2012measuring}. We capture GPU kernel execution time that excludes data initialization/transfer time. We use the \texttt{nvml} API~\cite{NVIDIAMa14} to measure GPU energy consumption.
We use CACTI \gfcrii{7.0}~\cite{cacti} to evaluate \prop and SIMDRAM energy consumption, where we take into account that each additional simultaneous row activation increases energy consumption by 22\%~\cite{seshadri2017ambit, hajinazarsimdram}.
\gfisca{We evaluate two SIMDRAM configurations:
\li~SIMDRAM with \gls{SLP}~\omcrii{\cite{kim2012case}} \gfasplos{and static bit-precision} (\emph{SIMDRAM\gfasplos{-SP}}), and 
\lii~SIMDRAM with \gls{SLP} and \prop' \emph{Dynamic Bit-Precision Engine} (\emph{SIMDRAM\gfasplos{-DP}})\gfmicro{. \gfasplos{In both configurations, the system implements only the 16 \uprogs proposed in SIMDRAM (i.e., there is \emph{no} \uproglib enabled). }
We evaluate four \prop configurations:}
\li~\prop \emph{LT\gfasplos{-SP}} and
\lii~\prop \emph{EN\gfasplos{-SP}}, where \prop \revCommon{selects} the \emph{lowest latency} \omcrii{(LT)} and \emph{lowest energy} \omcrii{(EN)} consuming \uprog, respectively, \agymicro{using the statically profiled bit-precision from Fig.~\ref{fig:narrow_values}};
\agymicro{\liii~\prop \emph{LT\gfasplos{-DP}} and \liv~\prop \emph{EN\gfasplos{-}DP}, where \prop executes the \emph{lowest latency} and \emph{lowest energy} consuming \uprog with dynamically chosen bit-precision.}\revdel{\agymicro{\footnote{\agymicro{For statically profiled bit-precision, we \emph{must} round the bit-precision up to the nearest power-of-two\revdel{, e.g., a 13-bit operation will be implemented using 16-bit data, since high-level programming language (e.g., C/C++) are \emph{bounded} by 2's complement data representation formats}.}}}}
We use 64 subarrays in \emph{\omcrii{only} one} DRAM bank for \omcrii{our} \gls{PuD} \omcrii{evaluations}.}}\footnote{\label{ft.rc.1}\revC{\gfmicro{The column/address (C/A) bus allows the simultaneously activation of up to 84 DRAM subarrays ($\frac{t_{RAS}}{t_{CK}}$ = $\frac{32~ns}{0.38~ns}$ = 84).}}}


\begin{table}[!ht]
   \caption{Evaluated system configurations.}
   \centering
   \footnotesize
   \tempcommand{1}
   \renewcommand{\arraystretch}{0.9}
   \resizebox{\columnwidth}{!}{
   \begin{tabular}{@{} c l @{}}
   \toprule
   \multirow{5}{*}{\shortstack{\textbf{Intel}\\ \textbf{Comet Lake CPU~\cite{intelcometlake}} \\ \omcrii{\textbf{(Real System)}}}} & x86~\cite{guide2016intel}, 16~cores, 8-wide, out-of-order, 3.8~GHz;  \\
                                                                           & \emph{L1 Data + Inst. Private Cache:} 256~kB, 8-way, 64~B line; \\
                                                                           & \emph{L2 Private Cache:} 2~MB, 4-way, 64~B line; \\
                                                                           & \emph{L3 Shared Cache:} 16~MB, 16-way, 64~B line; \\
                                                                           & \emph{Main Memory:} 64~GB DDR4-2133, 4~channels, 4~ranks \\
   \midrule
      \multirow{3}{*}{\shortstack{\textbf{NVIDIA}\\ \textbf{A100 GPU~\mbox{\cite{a100}}} \\ \omcrii{\textbf{(Real System)}}}} &  7~nm technology node; 826~mm$^2$ die area~\cite{a100}; 6912 CUDA cores;\\ 
                                                                            & \asplosrev{\hl{432 tensor cores}}, 108 streaming multiprocessors, 1.4~GHz base clock; \\
                                                                            & \emph{L2 Cache:} 40~MB L2 Cache; \emph{Main Memory:} 40~GB HBM2~\mbox{\cite{HBM,lee2016simultaneous}} \\
   \midrule

   \multirow{9}{*}{\shortstack{\textbf{SIMDRAM~\cite{hajinazarsimdram}}\\ \textbf{\& \prop} \\ \omcrii{\textbf{(Simulated)}}}} &  gem5-based in-house simulator~\omcrii{\cite{proteusgit, mimdramgit}};  x86~\cite{guide2016intel};  \\ 
                                                            & 1 \gfmicro{out-of-order core @ 4~GHz (\emph{only} for instruction offloading});\\
                                                                             & \emph{L1 Data + Inst. Cache:} 32~kB, 8-way, 64~B line;\\
                                                                             & \emph{L2 Cache:} 256~kB, 4-way, 64~B line; \\
                                                                             & \emph{Memory Controller:}  8~kB row size, FR-FCFS~\cite{mutlu2007stall,zuravleff1997controller}\\
                                                                             & \emph{Main Memory:}  \gfmicro{DDR5-5200}~\omcrii{\cite{jedec2020jesd795}}, 1~channel, 1~rank, 16~banks \\ 
                                      
   \bottomrule
   \end{tabular}
   }
   \label{table_parameters}
\end{table}

\label{r3.2A}\Copy{R3.2A}{\paratitle{Real-World Applications} We select \gfisca{twelve} workloads from four popular benchmark suites in our real-workload analysis (\gfisca{as Table~\ref{table:workload:properties} describes)}\revdel{, including 
\li~525.x264\_r (\texttt{x264}) from SPEC 2017~\cite{spec2017};
\lii~\texttt{pca} from Phoenix~\cite{yoo_iiswc2009};
\liii~\texttt{2mm}, 
\texttt{3mm}, 
convolution (\texttt{cov}),
doitgen (\texttt{dg}), 
fdt\gfmicro{d}-apml (\texttt{fdt\gfmicro{d}}),
gemm  (\texttt{gmm}), and
gramschmidt (\texttt{gs}) from Polybench~\cite{pouchet2012polybench};
and
\liv~heartwall (\texttt{hw}), kmeans (\texttt{km}), and backprop (\texttt{bp}) from Rodinia~\cite{che_iiswc2009}}.
We manually modified each workload to 
\li~identify loops that can benefit from \gls{PuD} computation, i.e., loops that are memory-bound and that can leverage \gls{SIMD} parallelism and
\lii~use the appropriate \emph{bbop} instructions.
\asplosrev{\hl{To identify loops that }\omrev{\hl{can leverage}}\hl{ \mbox{\gls{SIMD}} parallelism, we \omcriv{use the} \gfcrii{MIMDRAM compiler~\omcriv{\cite{mimdramgit}} for identification and generation} \omcriv{of \gls{PuD} instructions}, which uses LLVM's loop auto-vectorization engine~\mbox{\cite{sarda2015llvm, lopes2014getting,writingpass,lattner2008llvm}} as a profiling tool that outputs \mbox{\gls{SIMD}}-safe loops in an application. 
We use the clang compiler~\mbox{\cite{lattner2008llvm}} to compile each application while enabling the loop auto-vectorization engine and its loop vectorization report (i.e., \texttt{-O3 -Rpass-analysis=loop-vectorize -Rpass=loop-vectorize}).
We observe that applications with regular and wide data parallelism (e.g., applications operating over large dense vectors) are better suited for SIMD-based \mbox{\gls{PuD}} systems. 
We select applications from various domains, including linear algebra and stencil computing (i.e., 2mm, 3mm, doitgen, fdtd-apml, gemm, gramschmidt from Polybench~\mbox{\cite{pouchet2012polybench}}), machine learning (i.e., pca from Phoenix~\mbox{\cite{yoo_iiswc2009}}, covariance from Polybench~\mbox{\cite{pouchet2012polybench}}, kmeans and backprop from Rodinia~\mbox{\cite{che_iiswc2009}}), and image/video processing (i.e., heartwall from Rodinia~\mbox{\cite{che_iiswc2009}} and 525.x264\_r from SPEC 2017~\mbox{\cite{spec2017}}).}}}\revdel{\footnote{\label{r3.2B}\Copy{R3.2B}{\asplosrev{\hl{Note that \mbox{\gls{PuD}} architectures are not yet general-purpose solutions that are }\omrev{\hl{(easily)}}\hl{ well-suited for a }\omrev{\hl{very wide variety}}\hl{ of workloads. The authors of~\mbox{\cite{mimdramextended}} have conducted an extensive workload characterization analysis of 117 different applications from seven benchmark suites to identify kernels that are suited to \mbox{\gls{PuD}} architectures, with results on the twelve same applications we utilize in our paper.}}}}}\revdel{\footnote{Several prior works~\cite{damov,devic2022pim,dualitycache,fujiki2018memory,vadivel2020tdo,iskandar2023ndp,pattnaik2016scheduling} have previously shown that our selected 12 can benefit from different types of \gls{PIM} architectures.}}
\gf{Since our \omcrii{baseline} \gls{PuD} substrate (SIMDRAM) does \emph{not} support floating-point, we manually modify the selected floating-point-heavy \gls{PuD}-friendly loops to operate on fixed-point data arrays.\footnote{We only modify the three applications from Rodinia to use fixed-point, \sgi{as done by prior works~\cite{fujiki2018memory,yazdanbakhsh2016axbench,ho2017efficient}}. \sgi{Polybench} can be configured to use integers; the \gls{PuD}-friendly loops in \texttt{x264} and \texttt{pca} use integers.%
\sgdel{Prior works~\cite{fujiki2018memory,yazdanbakhsh2016axbench,ho2017efficient} also employ fixed-point for the same three Rodinia applications.}} We do \emph{not} observe an output quality degradation when employing fixed-point.}
{We use the largest input dataset available \gfcrii{\omcriv{for} each benchmark}.}

\label{r3.0}
\begin{table}[ht]
   \caption{Evaluated applications. \gfcriv{We measure peak GPU utilization and total memory footprint on a real system.}}
   \tempcommand{0.8}
   \resizebox{\columnwidth}{!}{%
   \Copy{R3.0}{
    \begin{tabular}{|c|c||c|c|c|c|}
\hline
\textbf{\begin{tabular}[c]{@{}c@{}}Benchmark\\  Suite\end{tabular}} & \textbf{\begin{tabular}[c]{@{}c@{}}Application\\ (Short Name)\end{tabular}} & \textbf{\begin{tabular}[c]{@{}c@{}}\revA{Peak GPU} \\ \revA{Util. (\%)}\end{tabular}} & \textbf{\begin{tabular}[c]{@{}c@{}} \asplosrev{\hl{Total Mem.}} \\ \asplosrev{\hl{Footprint (GB)}} \end{tabular}} & \textbf{\begin{tabular}[c]{@{}c@{}}Bit-Precision\\  \{min, max\}\end{tabular}} & \textbf{\begin{tabular}[c]{@{}c@{}}PUD \\ Instrs.\sgi{$^\dag$}\end{tabular}} \\ \hline \hline
\begin{tabular}[c]{@{}c@{}}Phoenix~\cite{yoo_iiswc2009}\end{tabular} & pca (\texttt{pca}) & \revA{--} & \asplosrev{\hl{1.91}} & \{8, 8\} & D, S, M, R \\ \hline
\multirow{7}{*}{\begin{tabular}[c]{@{}c@{}}Polybench\\ \cite{pouchet2012polybench}\end{tabular}} & 2mm (\texttt{2mm}) & \begin{tabular}[c]{@{}c@{}} \revA{98} \end{tabular} & \asplosrev{\hl{4.77}} & \{13, 25\} & M, R \\ \cline{2-6} 
 & 3mm (\texttt{3mm}) & \begin{tabular}[c]{@{}c@{}}\revA{100}\end{tabular} & \asplosrev{\hl{26.7}} & \{12, 12\} & M, R \\ \cline{2-6} 
 & covariance (\texttt{cov}) & \begin{tabular}[c]{@{}c@{}}\revA{100}\end{tabular} & \asplosrev{\hl{7.63}} & \{23, 23\} & D, S, R \\ \cline{2-6} 
 & doitgen (\texttt{dg}) & \begin{tabular}[c]{@{}c@{}}\revA{92}\end{tabular} & \asplosrev{\hl{33.08}} & \{10, 11\} & M, C, R \\ \cline{2-6} 
 & fdt\gfmicro{d}-apml (\texttt{fdt\gfmicro{d}}) & \begin{tabular}[c]{@{}c@{}}\revA{--}\end{tabular} & \asplosrev{\hl{36.01}}  & \{11, 13\} & D, M, S, A  \\ \cline{2-6} 
 & gemm (\texttt{gmm}) & \begin{tabular}[c]{@{}c@{}}\revA{98}\end{tabular} & \asplosrev{\hl{22.89}} & \{12, 24\} & M, R  \\ \cline{2-6} 
 & gramschmidt (\texttt{gs}) & \begin{tabular}[c]{@{}c@{}}\revA{66}\end{tabular} & \asplosrev{\hl{22.89}} & \{12, 13\} & M, D, R  \\ \hline
\multirow{3}{*}{\begin{tabular}[c]{@{}c@{}}Rodinia\\ \cite{che_iiswc2009}\end{tabular}} & backprop (\texttt{\revCommon{bp}}) & \revA{--}  & \asplosrev{\hl{22.50}} & \{13, 13\} & M, R  \\ \cline{2-6} 
 & heartwall (\texttt{hw}) &  \revA{48} & \asplosrev{\hl{0.03}} & \{17, 17\} & M, R  \\ \cline{2-6} 
 & kmeans (\texttt{km}) & \revA{36} & \asplosrev{\hl{1.23}} & \{17, 17\} & S, M, R  \\ \hline
\begin{tabular}[c]{@{}c@{}}SPEC 2017\\\cite{spec2017}\end{tabular} & 525.x264\_r (\texttt{x264}) & \begin{tabular}[c]{@{}c@{}}\revA{--}\end{tabular} & \asplosrev{\hl{0.15}} & \{1, 8\} & A, R  \\ \hline
\end{tabular}%
}
}
\resizebox{\columnwidth}{!}{$^\dag$ D = division, S = subtraction, M = multiplication, A = addition, R = reduction, C = copy}
    \label{table:workload:properties}
\end{table}

\section{Evaluation}
\label{sec:eval}

\subsection{Real-World Application Analysis} 
\label{sec:eval:real}

\begin{figure}[ht]
    \centering
   \includegraphics[width=\linewidth]{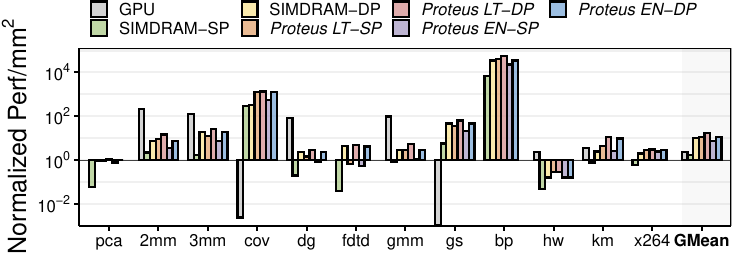}    \caption{\gfmicro{\revA{\gf{CPU-normalized performance per mm$^2$ for \gfisca{twelve} real-world applications\revdel{ in a CPU, GPU, SIMDRAM, \gfisca{SIMDRAM with dynamic bit-precision (\emph{SIMDRAM w/ DP}),} \prop latency-optimized (\prop \emph{LT}), and \prop energy-optimized (\prop \revA{\emph{EN}})}.
  Phoenix~\cite{yoo_iiswc2009} and SPEC2017~\cite{spec2017} do \emph{not} provide GPU implementations of \texttt{pca} and \texttt{x264}.}}}}
    \label{fig:real_workloads}
\end{figure}

\paratitle{\gfisca{Performance}} \gf{Fig.~\ref{fig:real_workloads} shows the CPU, GPU, SIMDRAM, \gf{and \prop}\revdel{SIMDRAM with dynamic bit-precision enabled (\emph{SIMDRAM w/ DP}), \prop \gfisca{latency-optimized (\prop \emph{LT}), and
\prop energy-optimized (\prop \emph{EN})}} performance for \gfisca{twelve} real-world applications. 
As in prior works~\revD{\cite{li2017drisa,lee20223d,zhou2023p,ferreira2021pluto,ferreira2022pluto}}, we report area-normalized results (i.e., performance per mm$^2$) for a fair comparison.\revdel{\footnote{\label{ft.rd.8}\revD{We report area-normalized performance results for fairness: the area occupied by \gls{PuD} systems (i.e., a single DRAM bank) is much smaller than the area occupied by processor-centric systems (i.e., a GPU and a CPU). The performance per $mm^2$ metric allows us to easier compare results across very distinct architectures while providing an indication of how the proposed system would scale (performance-wise) if more resources (e.g., more DRAM banks in a DRAM chip) were used for computing.}~\prtagD{D8}}} 
We make \gfmicro{four} observations. 
First, \prop \emph{significantly} outperforms all three baseline systems.  
On average across all \gfisca{twelve} applications, \gfisca{\prop \emph{LT\gfmicro{-DP}} (\prop \emph{EN\gfmicro{-DP}})} achieves 
17$\times$ (11.2$\times$), 
7.3$\times$ (4.8$\times$), and 
10.2$\times$ (6.8$\times$) the performance per mm$^2$ of the CPU, GPU, and SIMDRAM, respectively. 
Second, \gfisca{we observe that equipping SIMDRAM with \prop' \emph{Dynamic Bit-Precision Engine} to leverage narrow values for \gls{PuD} execution \emph{significantly} improves overall performance. On average\revdel{across all applications}, \emph{SIMDRAM-DP} provides 6.3$\times$ the performance \omcrii{per mm$^2$} of \emph{SIMDRAM-SP}. 
Third, \prop' ability to adapt the \uprog  depending on the target bit-precision further improves overall performance by 1.6$\times$ that of \emph{SIMDRAM-DP}.} 
\agymicro{Fourth, \prop' \dynengine \gfmicro{further increases performance by} \gfmicro{46\%}, \omcrii{over} \prop with static \gfmicro{bit-precision}.} This happens because for statically profiled bit-precision, we \emph{must} round the bit-precision up to the nearest power-of-two, \gfcrii{as high-level programming languages (e.g., C/C++) are \emph{inherently} constrained by the two's complement data representation}.
\revdel{We conclude that \prop significantly improves performance of bit-serial \gls{PuD} substrates.}}

\begin{figure}[ht]
    \centering
   \includegraphics[width=\linewidth]{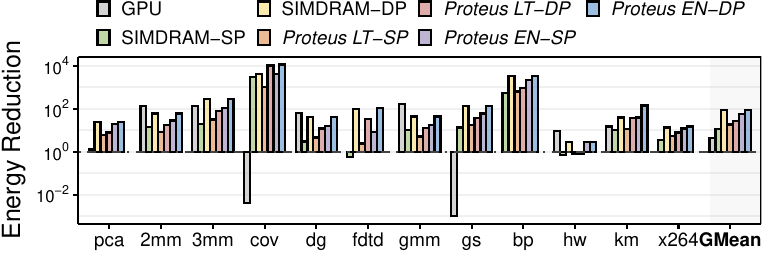}    \caption{\revA{\gf{\gfisca{\gfcrii{End-to-e}nd energy \gfcrii{reduction compared to the baseline CPU} for \gfisca{twelve} applications.\revdel{Values are normalized to the baseline CPU.}}}}}
    \label{fig:real_workloads:energy}
\end{figure}

\paratitle{\gfisca{Energy}} \gfisca{Fig.~\ref{fig:real_workloads:energy} shows the \gfcri{end-to-end energy reduction the} GPU, SIMDRAM, and \prop \gfcrii{provide compared to the baseline CPU} for twelve applications. We make \gfmicro{four} observations.
First, \prop \emph{significantly} reduces energy consumption compared to all three baseline systems.  
On average across all \gfisca{twelve} applications, \gfisca{\prop \emph{EN-DP} (\prop \emph{LT-DP})} \omcrii{provides}
90.3$\times$ (27$\times$), 
21$\times$ (6.3$\times$), and 
8.1$\times$ (2.5$\times$) \omcrii{lower energy consumption than} CPU, GPU, and \emph{SIMDRAM-SP}, respectively. 
Second, \gfcrii{enabling} \prop' \emph{Dynamic Bit-Precision Engine} and \uproglib \gfcrii{allows \prop to reduce energy consumption} by \sgi{an average of} 8$\times$ and 1.02$\times$ compared to \gfcrii{\gls{PuD} substrates with statically-defined bit-precision (}\sgi{\emph{SIMDRAM-SP}) and \gfcrii{bit-serial \emph{only} arithmetic (}\emph{SIMDRAM-DP}), respectively}. 
Third, compared to \emph{SIMDRAM-DP}, \prop \emph{LT-DP} \emph{increases} energy consumption by 3.3$\times$, on average. \revC{\label{rc.6}\changeC{C6}This is because the \omcrii{highest} performance implementation of a \gls{PuD} operation often leads to an increase in the number of \aaps required for \gls{PuD} computing. 
In many cases, the energy associated with inter-subarray data copies (employed in \gls{RBR} and bit-parallel algorithms) leads to an \emph{increase} in energy consumption. Even though the inter-subarray data copy latency can be hidden by leveraging \prop' \gls{OBPS} data mapping, the extra power the DRAM subsystem requires to perform them impacts overall energy consumption.}
\agymicro{Fourth, the \dynengine \gfmicro{further reduces} \prop' energy consumption by \gfmicro{58}\%, compared to \prop with static bit-precision.}
\revdel{We conclude that \prop is an energy-efficient \gls{PuD} substrate.} }

\subsection{\gfisca{Data \gfcrii{Mapping and Representation Format} Conversion Overheads}}
\label{sec:eval:dataconverson}

\begin{figure}[ht]
    \centering
   \includegraphics[width=0.85\linewidth]{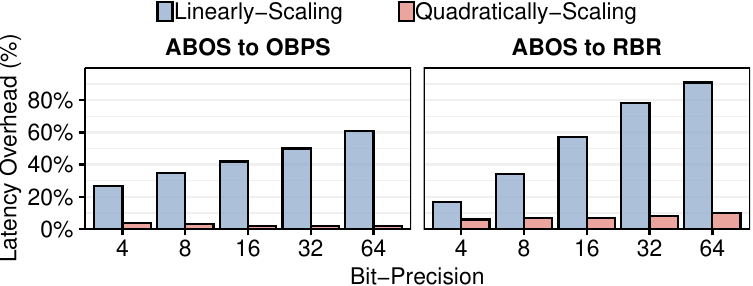}    
   \caption{\gfisca{\omcriv{Latency overheads of d}ata \gfcrii{mapping and representation} format conversion.\revdel{ \emph{ABOS} stands for all-bits in one-subarray; \emph{OBPS} stands for one-bit per-subarray; \emph{RBR} stands for redundant binary representation.}}}
    \label{fig:layout_conversion}
\end{figure}

\gfcrii{We evaluate the worst-case latency associated with 
\li~data \gfcrii{mapping conversion (from the conventional \gls{ABOS} data mapping to our \gls{OBPS} data mapping) and 
\lii~data representation} format conversion \gfcrii{(from \omcriv{ABOS} to \gls{RBR}) that \prop might perform during the execution of a \gls{PuD} operation.} 
Fig.~\ref{fig:layout_conversion} shows \gfmicro{the} worst-case data \gfcrii{mapping and representation} format conversion latency \gfcrii{overhead} for linearly- and quadratically-scaling \uprogs.}
We make two observations.

First, data \gfcrii{mapping and representation} format conversion can \emph{significantly} impact linearly-scaling \uprogs, \gfcrii{causing} up to 60\% and 91\% \gfcrii{latency} overhead when converting from \gls{ABOS} to \gls{OBPS} and from \gls{ABOS} to \gls{RBR}, respectively. 
Second, in contrast, the impact of data \gfcrii{mapping and representation} format conversion in quadratically\omcrii{-}scaling \uprogs is low (i.e., less than 10\% latency overhead). 
This difference is because the number of in-DRAM operations required to perform data \gfcrii{mapping and representation} format conversion increases \emph{linearly} with the bit-precision (\cref{sec:implementation:alltogether}).
In most cases, the data \gfcrii{mapping and representation} format conversion is a one-time overhead that an application pays when executing a series of \gls{PuD} operations.  
On average, across our 12 applications, data \gfcrii{mapping and representation} format conversion accounts for 7.2\% of the total execution time.

\subsection{\asplosrev{Performance of Floating-Point Operations}}
\label{sec:eval:fp}

\label{r1.2A}\Copy{R1.2A}{\asplosrev{\hl{
We evaluate \prop' throughput for floating-point arithmetic operations.
Since our underlying \mbox{\gls{PuD}} architecture, i.e., SIMDRAM, does \emph{not} natively support floating-point operations, we demonstrate how \mbox{\prop} can be leveraged in a different \mbox{\gls{PuD}} architecture that can be modified to support floating-point data formats, i.e., DRISA~\mbox{\cite{li2017drisa}}. 
The main limitation of SIMDRAM when supporting floating-point computation is the lack of interconnects across DRAM columns, which is required for exponent normalization during floating-point addition. 
Since the DRISA architecture includes a shifting network within a DRAM subarray, }\omrev{\hl{it}}\hl{ can perform the required exponent normalization.}}}
\label{r1.2B}\Copy{R1.2B}{\asplosrev{\hl{We perform a synthetic throughput analysis of in-DRAM vector addition/subtraction and multiplication/division operations.}}}\footnote{\label{r1.2C}\Copy{R1.2C}{\asplosrev{\hl{We evaluate synthetic workloads instead of our twelve real-world applications since there is \emph{no} publicly available tool-chain to map real-world applications to }\omrev{\hl{the}}\hl{ baseline DRISA~\mbox{\cite{li2017drisa}} architecture. A similar approach is \omcrii{followed} in~\mbox{\cite{hajinazarsimdram}}.}}}} \label{r1.2D}\Copy{R1.2D}{\asplosrev{\hl{We use 64M-element input arrays, with randomly initialized single-precision floating-point data. We }\omrev{\hl{evaluate}}\hl{ two system configurations:
\li~DRISA 3T1C~\mbox{\cite{li2017drisa}} architecture, which performs \emph{in-situ} \texttt{NOR} }\omrev{\hl{computation}}\hl{; and
\lii~DRISA 3T1C architecture coupled with \mbox{\prop}. 
We make two observations.
First, we observe that the DRISA 3T1C architecture coupled with \mbox{\prop} achieves 1.17$\times$}\omrev{\hl{/1.15$\times$}}\hl{ and 1.38$\times$}\omrev{\hl{/1.37$\times$}}\hl{ the arithmetic throughput of the baseline DRISA 3T1C for floating-point addition/subtraction and multiplication/division operations, respectively.
Second, \prop' throughput gains are more prominent for  multiplication/division operations, since it can \omcrii{reduce} costly in-memory multiplication/division operations during mantissa computation by leveraging narrow mantissa values. 
We conclude that \prop' key ideas apply to different underlying in-DRAM processing techniques and data \omcrii{types}.}}}

\subsection{\asplosrev{\prop vs.\ Tensor Cores \omrev{in} GPU\omrev{s}}}
\label{sec:eval:tensor}

We compare the performance and energy efficiency of our real-world applications \gfcriv{that perform \gls{GEMM} operations} while running on the tensor cores in the NVIDIA A100 GPU and \mbox{\prop} for narrow data precision input operands (i.e. 4-bit and 8-bit integers).
To do so, we 
\li~identify the subset of our real-world applications that mainly perform \mbox{{GEMM}} operations and therefore are suitable for the A100's tensor core engines; and
\lii~re-implement such workloads using optimized instructions (from NVIDIA's CUTLASS~\mbox{\cite{cutlass}}) to perform tensor \mbox{{GEMM}} operations on the A100 GPU tensor cores. 
Re-implementing the GPU workloads is necessary since GPU tensor core instructions are \emph{not} automatically produced via the standard CUDA code our workloads use and there is \emph{no} reference implementation available from the original benchmark suites targeting tensor core GPUs.
\revdel{We re-implement three workloads from our real-world applications, i.e., \texttt{2mm}, \texttt{3mm}, and \texttt{gmm}, for tensor core execution.
To ensure that our re-implemented workloads efficiently utilize the tensor cores, we leverage NVIDIA's open-source CUTLASS library~\mbox{\cite{cutlass}}, which provides CUDA C++ template abstractions for high-performance \mbox{{GEMM}} operations and C++ APIs for non-standardized data types (4-bit integers, in our case). During data initialization, we ensure that the input data fits into 4-bit or 8-bit data types, depending on the evaluated configuration. As in~\mbox{\cref{sec:eval:real}}, we capture the execution time of the GPU kernel excluding data initialization/transfer time, and we use the \texttt{nvml} API~\mbox{\cite{NVIDIAMa14}} to measure GPU energy consumption.} We employ \omcrii{A100's} \omcriv{all} 432 tensor cores during GPU execution.

\begin{figure}[ht]
    \centering
    \Copy{R2.3}{
   \includegraphics[width=\linewidth]{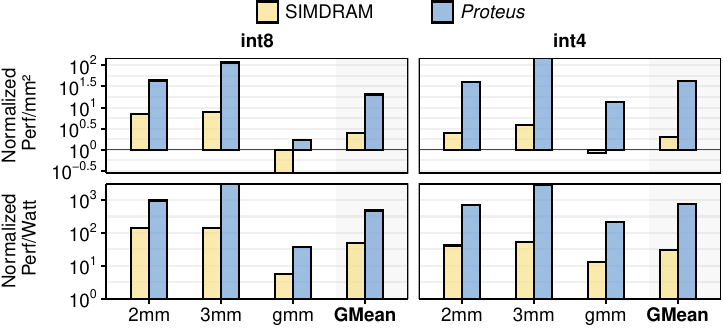}    
   \caption{\gfcrii{\hl{Performance per mm$^2$~(top) and performance per Watt~(bottom) of GEMM-intensive real-world applications using \mbox{\texttt{int8}} and \mbox{\texttt{int4}}}\omrev{\hl{, normalized to the same metric measured on 432 NVIDIA A100 tensor cores.}}}}
   \label{fig:real_workloads:tensorgpu}
}
\end{figure}

\label{r2.2}\Copy{R2.2}{\asplosrev{\hl{Fig.~\mbox{\ref{fig:real_workloads:tensorgpu}} shows the tensor cores, SIMDRAM, and \mbox{\prop} performance per mm$^2$ (Fig.~\mbox{\ref{fig:real_workloads:tensorgpu}}\gfcrii{, top}) and energy efficiency (i.e., performance per Watt in Fig.~\mbox{\ref{fig:real_workloads:tensorgpu}}\gfcrii{, bottom}) for three GEMM-heavy real-world applications using 8-bit (\texttt{int8}) and 4-bit (\texttt{int4}) data types. Values are normalized to \omcrii{those obtained on real} GPU tensor cores. We make two observations.
First, \mbox{\prop} significantly improves performance }\omrev{\hl{per mm$^2$}}\hl{ and energy efficiency compared to both \omcrii{tensor cores and SIMDRAM} across all applications and data types. 
}\omrev{\hl{On average across the three applications, \mbox{\prop} \omcrii{provides} 
\li~20$\times$/43$\times$ and 8$\times$/21$\times$ the performance per mm$^2$ and 
\lii~484$\times$/767$\times$ and 9.8$\times$/25$\times$ the performance per Watt of the tensor cores and SIMDRAM, respectively, using \texttt{int8}/\texttt{int4} data types.}}\hl{  
\mbox{\prop} and SIMDRAM are capable of outperforming the tensor cores of the A100 GPU for narrow data precisions since }\omrev{\hl{both}} the throughput and the energy efficiency of bit-serial \mbox{\gls{PuD}} architectures \emph{increase} quadratically for multiplication operations as the bit-precision \emph{decreases}~\mbox{\cite{hajinazarsimdram}}. 
\revdel{As a comparison point, both SIMDRAM and \prop fail to outperform the baseline CUDA cores in the A100 GPU for the same three workloads when computing over a higher dynamic range (see \cref{sec:eval:area}). However, such performance and energy efficiency gaps shift in favor of SIMDRAM and \prop, particularly when we move from 8 to 4 bit input  operands.}  
Second, we observe that by employing dynamic bit-precision and adaptive arithmetic computation, \mbox{\prop} further improves the performance and energy gains that SIMDRAM provides compared to the A100 GPU's tensor cores, even improving performance compared to the \omrev{\hl{tensor cores}}\hl{ in cases where SIMDRAM fails to do so (i.e., for \texttt{gmm}). 
}}}

\subsection{Area Analysis}
\label{sec:eval:area}

\paratitle{DRAM Chip Area \omcriv{and Storage} Overhead} \gf{We use CACTI 7.0~\cite{cacti} to evaluate the area overhead of the primary components in the \prop design using a 22~nm technology node. 
\prop does \emph{not} introduce any modifications to the DRAM array circuitry other than those proposed by 
\li~Ambit, which has an area overhead of $<$1\% in a commodity DRAM chip~\cite{seshadri2017ambit};
\lii~LISA, which has an area overhead of 0.6\% in a commodity DRAM chip~\cite{chang2016low}; and
\liii~SALP, which has an area overhead of \gfcrii{0.15}\% in a commodity DRAM chip~\cite{kim2012case}.} 
\revB{\label{rb.3}\changeB{B3}We reserve less than 1 DRAM row (i.e., \SI{6.25}{\kilo\byte} in an \SI{8}{\giga\byte}) to store our implemented \uprogs. 
\revdel{That is enough space to store all 16 SIMDRAM \uprogs, \prop' \uprogs for addition, multiplication, division, and subtraction for the different implemented algorithms and data mapping schemes.}
In total, \gfcrii{we implement} 50 \uprogs, each of which takes \SI{128}{\byte} of DRAM space. }

\paratitle{\gfmicro{CPU} Area Overhead} 
\gf{\revdel{The main components in the \prop \emph{Control Unit} are the 
\uproglib and \dynengine.} 
We size the \uproglib to contain:
\li~16 64~B \glspl{LUT}, each \gls{LUT} holding a 8-bit  \idx);
\lii~one 2~kB \emph{\uprog Scratchpad} Memory. The size of the \uproglib is enough to hold one \gls{LUT} per SIMDRAM \gls{PuD} operations and address $2^8$ different \uprog implementations. The size of the \uprog{} Scratchpad is large enough to store the \uprog{}s for all 16 SIMDRAM operations.
We use a 128~B scratchpad for the \dynengine. 
\gfcrii{Using CACTI,} we estimate that the \prop \emph{Control Unit} area is 0.03~mm$^2$.} 
%
\gfmicro{\prop' \emph{Data Transposition Unit} \asplosrev{\hl{(one per DRAM channel)}} uses}  an 8~kB fully-associative cache with a 128-bit cache line size for the \emph{Object Tracker}, and
two 4~kB transposition buffers.
\gfcrii{Using CACTI,} we estimate the \emph{Data Transposition Unit} area is 0.06~mm$^2$. 
Considering the area of the control and transposition units, \prop has an area overhead of only 0.03\% compared to the die area of an Intel Xeon E5-2697 v3 CPU~\cite{dualitycache}.  
\section{Related Work}

\gfcrii{To our knowledge, \prop is the first \omcriv{system that can} transparently execute \gls{PuD} operations with the \omcriv{best} bit-precision, data representation, and algorithm arithmetic implementation. 
We highlight \prop' key contributions by contrasting them with state-of-the-art \gls{PIM} designs.}

\paratitle{Processing-Using-DRAM} 
\gf{Prior works propose different ways of implementing \gls{PuD} operations~\cite{seshadri2013rowclone,seshadri2017ambit,xin2020elp2im,deng2018dracc,gao2019computedram,angizi2019graphide,hajinazarsimdram,li2018scope,ferreira2022pluto,deng2019lacc,kim2019d,li2017drisa,zhou2022flexidram, mimdramextended}\revdel{, either by 
\li~using the memory arrays themselves to perform  operations in bulk~\cite{seshadri.bookchapter17, seshadri2013rowclone,seshadri2018rowclone,chang2016low,wang2020figaro,seshadri2017ambit, xin2020elp2im, besta2021sisa,deng2018dracc, gao2019computedram,li2017drisa,angizi2019graphide, hajinazarsimdram,li2018scope,ferreira2021pluto,ferreira2022pluto,deng2019lacc,olgun2021quactrng,bostanci2022dr,kim2019d};
\lii~modifying the DRAM sense amplifier design with logic gates for computation~\cite{li2017drisa,zhou2022flexidram}}. 
\revdel{Even though such works leverage different algorithms and execution models to implement \gls{PuD} operations, they all assume a fixed bit-precision and algorithmic implementation. To improve performance,} Such works could benefit from \prop' dynamic bit-precision selection and alternative data representation and algorithms, since they all assume a static bit-precision and algorithmic implementation.
\gfisca{AritPIM~\cite{leitersdorf2023aritpim} provides a collection of bit-parallel and bit-serial algorithms for \gls{PuM} arithmetic. Compared to AritPIM, \prop
\li~\gfcrii{extends AritPIM's set} of bit-parallel algorithms for \omcriv{\gls{PuD}};
\lii~evaluates different data \gfcrii{mapping and format} \omcriv{representations} that lead to further performance and energy improvements; 
\liii~proposes a framework that can dynamically adapt to the bit-precision of the operation.}} 

\paratitle{Using Bit-Slicing Compilers \asplosrev{\hl{\& Early Termination}} for \gls{PIM}} \gf{Prior works~\cite{peng2023chopper,arora2023comefa} propose bit-slicing compilers for bit-serial \gls{PIM} computation. In particular, CHOPPER~\cite{peng2023chopper} improves SIMDRAM's programming model by leveraging bit-slicing compilers and employing optimizations to reduce the latency of a \uprog\revdel{, such as employing \gls{BLP} to overlap the latency of data copy and computation for \gls{PuD} operations targeting different DRAM banks}. Compared to CHOPPER, \prop has two main advantages. 
First, \prop improves \uprog performance by leveraging the DRAM parallelism within a single DRAM bank \gfcrii{via \gls{SLP}}.
Second, although bit-slicing compilers can naturally adapt to different bit-precision values, they require the programmer to specify the target bit-precision manually. 
In contrast, \prop \emph{dynamically} identifies the most suitable bit-precision transparently from the programmer.}
\asplosrev{\hl{Some other prior works (e.g.,~\mbox{\cite{caminal2022accelerating, wong2023pumice}}) propose techniques to realize \gfcrii{dynamic bit-precision-based} \mbox{\gls{PuM}} operations for different memory technologies. Compared to these, \prop' main novelty lies in realizing \gfcrii{dynamic bit-precision} of bit-serial \gfcrii{and bit-parallel} operation in the context of DRAM/majority-based \mbox{\gls{PuD}} systems.}}

\section{Conclusion}
\label{sec:conclusion}

\gfcrii{We introduce \prop, a data-aware hardware runtime framework that addresses the high execution latency of bulk bitwise \gls{PuD} operations.
To do so, \prop \emph{dynamically} \omcriv{adjusts} the bit-precision of \gls{PuD} operations by exploiting narrow values, and, based on that, \emph{chooses} and \emph{uses} the most appropriate data representation (i.e., two's complement or redundant-binary representation) and arithmetic algorithm implementation (i.e., bit-serial or bit-parallel) for \gls{PuD} systems. 
We demonstrate that \prop provides \omcriv{large performance and energy} benefits over state-of-the-art CPU, GPU, and \gls{PuD} systems. 
\gfcriv{The source code of \prop is freely available at \url{https://github.com/CMU-SAFARI/Proteus}.}
}

\section*{\gfcrii{Acknowledgments}}

\gfcrii{We thank the anonymous reviewers of ASPLOS 2024, ISCA 2024, MICRO 2024, ASPLOS 2025, and ICS 2025 for their feedback. 
We thank the SAFARI Research Group members for providing a stimulating intellectual environment. 
We acknowledge the generous gifts from our industrial partners, including Google, Huawei, Intel, and Microsoft. 
This work is supported in part by the ETH Future Computing Laboratory
(EFCL), Huawei ZRC Storage Team, Semiconductor Research Corporation, AI Chip Center for Emerging Smart Systems (ACCESS), sponsored by InnoHK funding, Hong Kong SAR, and European Union's Horizon programme for research and innovation [101047160 - BioPIM].}

\balance
{
  \bstctlcite{IEEEexample:BSTcontrol}
  \let\OLDthebibliography\thebibliography
  \renewcommand\thebibliography[1]{
    \OLDthebibliography{#1}
    \setlength{\parskip}{0pt}
    \setlength{\itemsep}{0pt}
  }
  \bibliographystyle{IEEEtran}
  \bibliography{refs}
}
    
\end{document}